\documentclass[english]{IEEEtran}
\usepackage[T1]{fontenc}
\usepackage[latin9]{inputenc}
\usepackage{graphicx}

\makeatletter
\usepackage{url}
\usepackage{booktabs} 

\makeatother

\usepackage{babel}
\begin{document}
\title{Scalability and Resilience of Software-Defined Networking: An Overview}
\author{
	\IEEEauthorblockN{Benjamin J. van Asten, Niels L. M. van Adrichem and Fernando A. Kuipers}\\
	\IEEEauthorblockA{Network Architectures and Services, Delft University of Technology\\
	Mekelweg 4, 2628 CD Delft, The Netherlands\\
	\{B.J.vanAsten@student., N.L.M.vanAdrichem@, F.A.Kuipers@\}tudelft.nl}
}
\maketitle{}
\begin{abstract}
Software-Defined Networking (SDN) allows to control the available
network resources by an intelligent and centralized authority in order
to optimize traffic flows in a flexible manner. However, centralized
control may face scalability issues when the network size or the number
of traffic flows increases. Also, a centralized controller may form
a single point of failure, thereby affecting the network resilience. 

This article provides an overview of SDN that focuses on (1) scalability
concerning the increased control overhead faced by a central controller,
and (2) resiliency in terms of protection against controller failure,
network topology failure and security in terms of malicious attacks.
\end{abstract}

\section{Introduction}

\label{sec:Introduction}

Currently, most switching and routing solutions integrate both data
and control plane functionality. The data plane performs per-packet
forwarding based on look-up tables located in the memory or buffer
of the switch or router, whereas the control plane is used to define
rules based on networking policies to create the look-up tables. Due
to the high demands on network performance and growing configuration
complexity, the control plane has become overly complicated, inflexible
and difficult to manage. To solve this problem a new networking paradigm
was needed, which was compatible with the widely used Ethernet switching
and IP routing techniques. The solution was found in virtualization
techniques used in server applications, where an abstraction layer
is positioned above the server hardware to allow multiple virtual
machines to share the available resources of the server. Software
Defined Networking (SDN) adopted this paradigm and introduced an abstraction
layer in networking.

By abstracting the network resources, the data and control planes
are separated. The data plane is located at the switch hardware, where
the optimized forwarding hardware is preserved and the control of
the network is centralized into an intelligent authority with the
aim to improve flexibility and manageability. A centralized authority
provides the intelligence to network switches to route and control
the traffic through the network infrastructure. Optimal paths through
the network can be provided by the central authority in advance or
on demand. The current implementation of the SDN networking paradigm
is found in the OpenFlow protocol developed by Stanford University
in 2008 and is currently under development within the Open Networking
Foundation. OpenFlow has attracted some big vendors in the networking
community and became the most popular realization of the SDN networking
paradigm.

Since the introduction of OpenFlow, much research has been performed
on two different fields, being i) scalability and performance of the
central authority in relation to the growth of network traffic and
requests, and ii) the robustness and resiliency of the network against
link and switch failures in the network, but also failures of the
central authority. Clearly, the two are related, since scalability
issues may cause failures.

In this overview, we specifically focus on scalability and resilience
of SDN. In section \ref{sec:RelatedWork}, we our work from existing
surveys on SDN and OpenFlow networking. We describe the basics behind
the SDN paradigm, the OpenFlow protocol, network controllers and compliant
switches in section \ref{sec:Software-Defined-Networking}. The general
framework and standard notation is given in section \ref{sec:OverviewFramework},
while sections \ref{sec:Scalability-in-SDN} to \ref{sec:Resiliency-in-SDN}
discuss related work in relation to our framework. Section \ref{sec:Conclusion}
concludes this overview.

\section{Related Surveys}

\label{sec:RelatedWork}

Nunes et al. \cite{mendoncca2013survey} presented a standard survey
with emphasis on past, present and future implementations of SDN.
It gives a proper overview of possible applications that could benefit
from SDN. Feamster et al. \cite{feamster2013road} give a historical
insight in the development of SDN networks, with the emphasis on virtualizing
the network and separating the data and control planes. A survey on
security in SDN is given by Scot-Hayward et al. \cite{scott2013sdn}.
The survey provides a nice categorization on security-related research
and addresses security analysis, enhancements and solutions, as well
as the data, control and application layer of SDN. Yeganeh et al.
\cite{yeganeh2013scalability} focus on the scalability concerns in
relation to current-state networking, controllers and switching hardware.
Sezer et al. \cite{sezer2013we} discus the implementation challenges
for SDN in relation to carrier-grade networks. Suzuki et al. \cite{suzuki2014survey}
take a similar approach concerning OpenFlow technologies in carrier-grade
and data center networks. In \cite{lara2013network}, Lara et al.
provide an extensive survey on network innovations using OpenFlow,
where the OpenFlow specification is discussed in detail and recent
experiences with OpenFlow deployments on campus networks and testbeds
are shared.

In contrast to the above mentioned surveys, we propose a graphical
reference framework, with which SDN strengths and frailties are identified
more easily. Furthermore, we specifically focus on scalability and
resilience aspects.

\section{Introduction to \protect \\
Software-Defined Networking}

\label{sec:Software-Defined-Networking}

In this section we introduce Software-Defined Networking (section
\ref{sub:Abstracting-the-network}), the OpenFlow protocol (section
\ref{sub:Description-of-OpenFlow}), OpenFlow controllers (section
\ref{sub:Description-of-OpenFlowController}) and Open vSwitch (section
\ref{sub:Description-of-OVS}).

\subsection{Abstracting the network}

\label{sub:Abstracting-the-network}

\begin{figure}
\includegraphics[width=1\columnwidth]{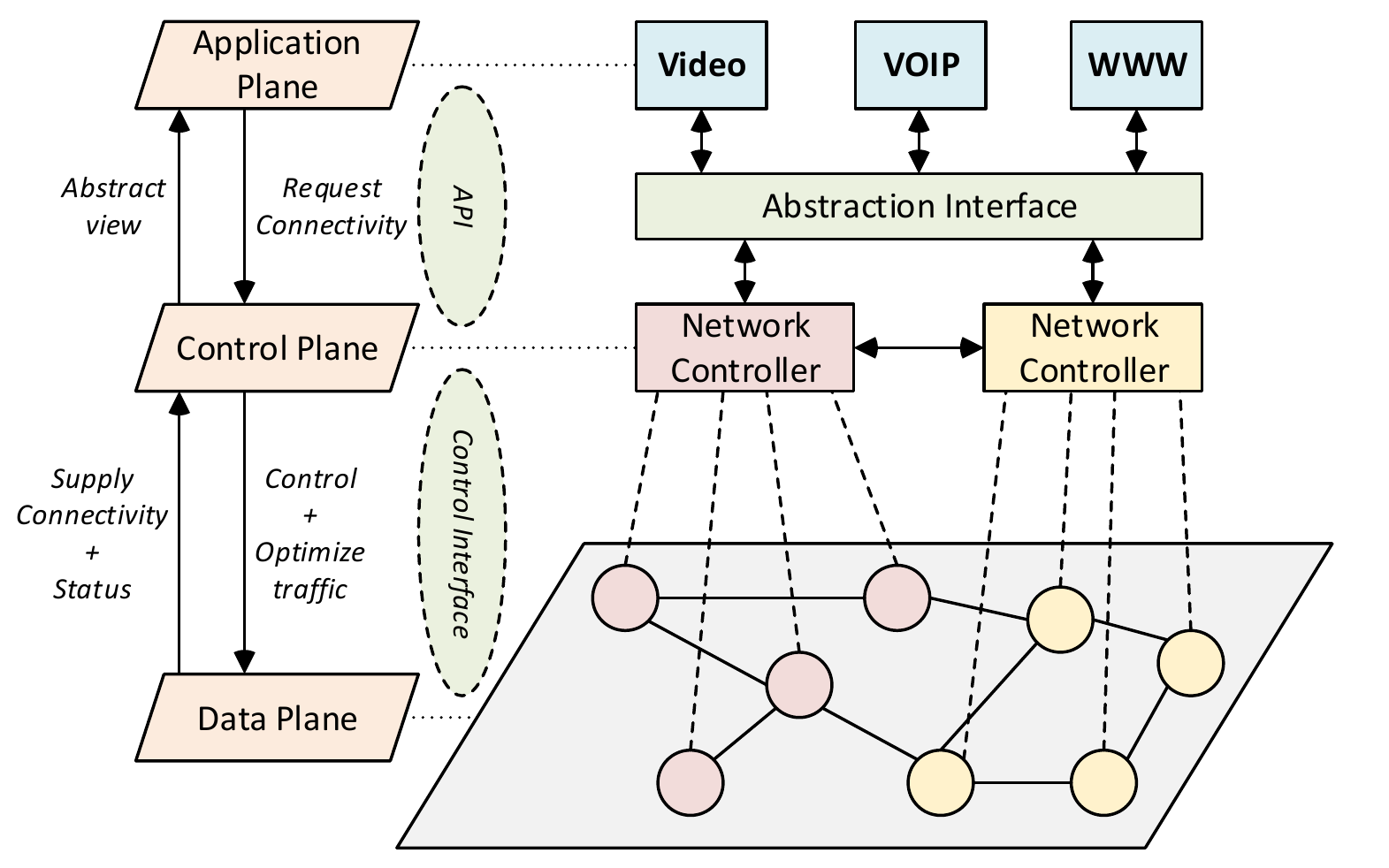}\protect\caption[SDN concept of abstracting the network view]{SDN concept of abstracting the network view - \emph{On the Data Plane,
network elements (switches) provide network connectivity and status
to the Control Plane. The network elements are configured by Network
Controllers via a Control Interface for global optimized configuration.
An abstract view of the network is given to the Application Plane
via a standardized interface. Network services request connectivity
from the Network Controllers, after which the Network Elements are
configured.}}

\label{fig:SDNConcept}
\end{figure}

In the SDN philosophy, the network topology is configured based on
requests from network services and applications. Services request
connectivity to a network and if the request can be fulfilled, paths
through the topology are provided to the service for the requested
amount of time. In figure \ref{fig:SDNConcept} the SDN concept is
presented.

The SDN concept speaks of three planes, which do not correspond directly
with the OSI reference model. A short description of the planes is
given below:
\begin{itemize}
\item \emph{Data Plane} - The Data Plane is built up from Network Elements
and provides connectivity. Network Elements consist of Ethernet switches,
routers and firewalls, with the difference that the control logic
does not make forwarding decisions autonomously on a local level.
Configuration of the Network Elements is provided via the control
interface with the Control Plane. To optimize network configuration,
status updates from the elements are sent to a Network Controller;
\item \emph{Control Plane} - Network Controllers configure the Network Elements
with forwarding rules based on the requested performance from the
applications and the network security policy. The controllers contain
the forwarding logic, but can be enhanced with additional routing
logic. Combined with actual status information from the Data Plane,
the Control Plane can compute optimized forwarding configurations.
To the application layer, an abstract view from the network is shared
via a general Application Programming Interface (API). This abstract
view does not contain details on individual links between elements,
but enough information for the applications to request and maintain
connectivity;
\item \emph{Application Plane} - Applications request connectivity between
two end-nodes, based on delay, throughput and availability descriptors
received in the abstract view from the Control Plane. The advantage
is the dynamic allocation of requests, as non-existing connectivity
does not need processing at local switch level. Also applications
can adapt service quality based on received statistics. For example
to throttle the bandwidth for video streaming applications on high
network utilization.
\end{itemize}
\begin{figure}
\includegraphics[width=1\columnwidth]{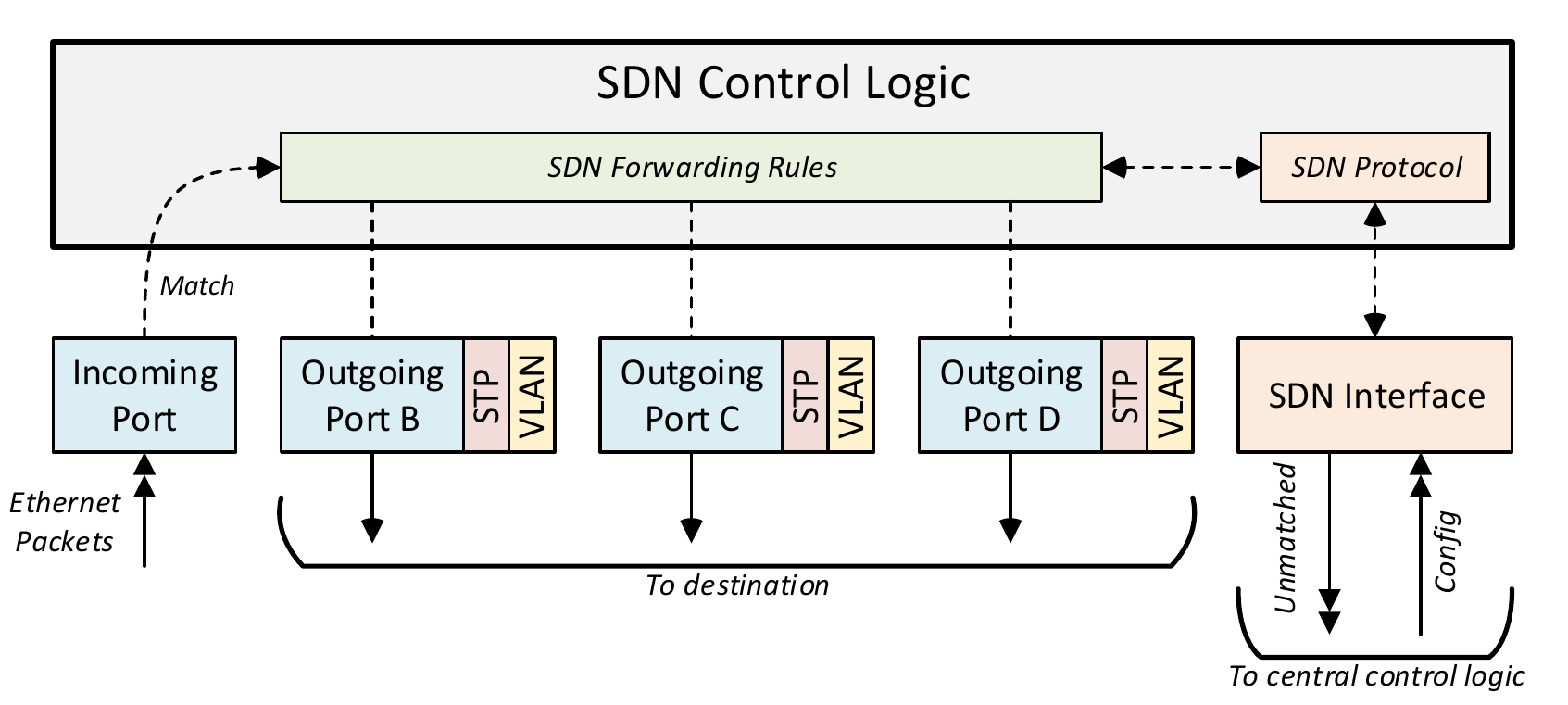}\protect\caption[SDN-enabled Ethernet switch]{SDN-enabled Ethernet switch - \emph{Incoming traffic is matched to
SDN Forwarding Rules and on positive matching traffic is forwarded
as normal. Information of unmatched packets is sent to the central
control logic (Network Controllers), where new SDN forwarding rules
are computed and configured at the SDN switches involved for transporting
the data packets. The local control logic is enhanced with a configuration
interface (SDN protocol) to communicate with Network Controllers.} }

\label{fig:SDNState}
\end{figure}

By decoupling the control logic the management of switches simplifies,
as decisions to flood or forward data packets are not made locally
anymore. Header information from data packets at the switches must
be transmitted to the central control logic for processing and configuration
computations, which introduces an additional delay in packet forwarding.
As seen in figure \ref{fig:SDNState}, the basic functionality of
the SDN switch is similar to that of an Ethernet switch. Header information
is matched to the configured SDN Forwarding Rules and packets are
subsequently forwarded to the configured outgoing port(s). Unmatched
header information is sent to the central control logic via the SDN
control interface. Thus, for communication between the SDN switch
and the centralized controllers an additional protocol is needed.
This protocol must contain the functionality to configure forwarding
rules and ports, as well be able to collect and transmit switch status
and statics to the central control logic. OpenFlow is such a protocol.

\subsection{OpenFlow protocol}

\label{sub:Description-of-OpenFlow}

\begin{figure}
\includegraphics[width=1\columnwidth]{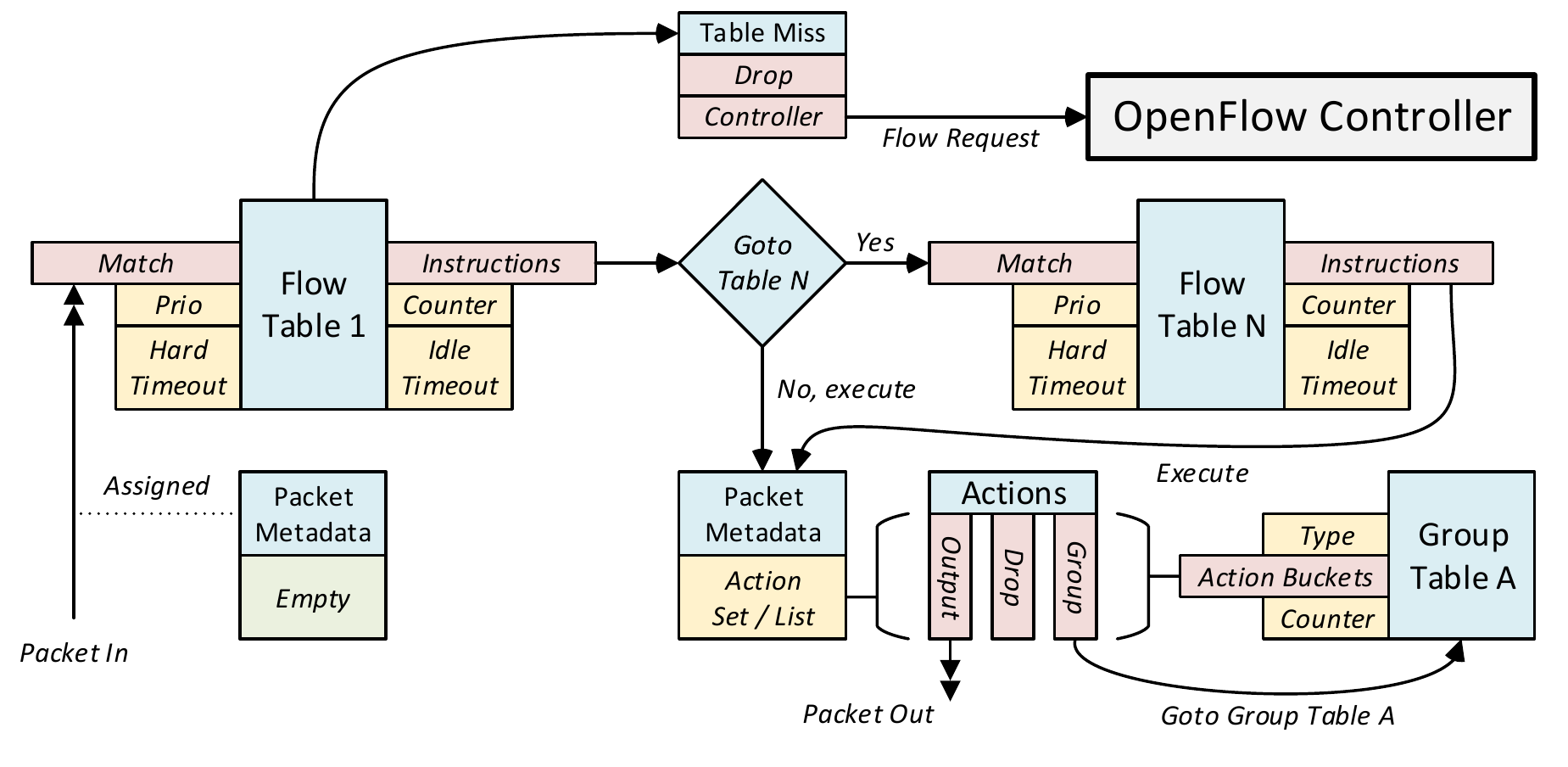}\protect\caption[Flow diagram of the OpenFlow protocol]{Flow diagram of the OpenFlow protocol - \emph{An incoming packet
is assigned with Packet Metadata and is matched to Flow Rules in Flow
Tables. Each Flow Rule contains instructions, which are added as Actions
to the Packet Metadata. Instructions can include forwarding to other
Flow Tables. If all Flow Tables are passed, the Actions in the Metadata
are executed. Actions define the outcome for the packets, such as
Output or Drop traffic. If a group of Flow Rules requires the same
processing, Group Tables are applied. Group tables also contain Actions.
When a packet does not match, a Table Miss is initiated and the packet
can be forwarded to the OpenFlow controller for further processing
or the packet is dropped.}}

\label{fig:OpenFlow-Description}
\end{figure}

Two examples of SDN protocols are OpenFlow \cite{ONF2013OpenFlowSpec}
and ForCES \cite{rfc5810}. More protocols exist, however OpenFlow
is most popular. OpenFlow has attracted many researchers, organizations
and foundations, so a wide collection of open-source software is available
in the form of OpenFlow controllers (section \ref{sub:Description-of-OpenFlowController}),
as well as physical and virtual switch implementations (section \ref{sub:Description-of-OVS}).
The OpenFlow protocol describes and couples switching hardware to
software configurations, such as incoming and outgoing ports, as well
as an implementation of the SDN Forwarding Rules. In figure \ref{fig:OpenFlow-Description}
a flow diagram is given of the matching process of an incoming packet
in an OpenFlow-compliant switch enabled with protocol version $1.3$.
A detailed survey on the OpenFlow specification is given in \cite{lara2013network}.

The SDN Forwarding Rules, called Flows in OpenFlow, are stored in
one or more Flow Tables. For each incoming packet, a metadata set
is created, containing an Action List, Action Set or both. In the
Action List and Set actions are added for each Flow Table the packet
transverses, whereas the Actions define the appropriate operations
for the packet. Examples of Actions are forward the packet to port
\emph{X}, drop the packet, go to Group Table \emph{A} or modify the
packet header. The main difference between a List and Set is the time
of execution. Actions added to a List are executed directly after
leaving the current Flow Table, whereas the Actions defined in the
Set are accumulated and executed when all Flow Tables are processed.
Each Flow Table contains Flow Entries with six parameters \cite{ONF2013OpenFlowSpec}: 
\begin{itemize}
\item \emph{Match} - The criteria to which the packets are matched. Criteria
include parameters of the datalink, network and transport layers contained
in data packet headers and optionally metadata from previous tables.
A selection of criteria is given in table \ref{tab: SelectionMatchField};
\item \emph{Instructions} - When a packet matches, instructions are added
to the metadata set to direct the packet to another Flow Table or
add Actions to the Action List or Set; 
\item \emph{Priority} - The packet header can match to multiple Flow Entries,
but the entry with highest priority determines the operations;
\item \emph{Counter} - Every time a packet has matched and is processed
by a Flow Entry, a counter is updated. Counter statistics can be used
by the OpenFlow controller and Application Plane to determine network
policies or for network monitoring \cite{vanopennetmon}; 
\item \emph{Hard Timeout} - A Flow Entry is added by an OpenFlow controller,
where the maximum amount of time this entry may exist in the Flow
Table before expiring is defined by the Hard Timeout. The Hard Timeout
can be used to limit network access for a certain node in the network
and for automatic refreshing of the Flow Table to prevent large tables;
\item \emph{Idle Timeout} - The amount of time a Flow Entry is not matched
is defined as the idle time. Idle Timeout defines the maximum idle
time and is mainly used for refreshing Flow Tables.
\end{itemize}
\begin{table}
\protect\caption{\label{tab: SelectionMatchField}Selection of fields for Flow Rules
to match incoming packets \cite{ONF2013OpenFlowSpec}. }
\centering
\begin{tabular}{l l l}

	\emph{Match Field} & \emph{Layer} & \emph{Description}  \\
\toprule[2pt]

	Ingress Port 	  & Physical & Incoming ports and interfaces   \\
\midrule[0.3pt]
	Ethernet Address	& Datalink & Source and destination MAC-address \\
\midrule[0.3pt]
	VLAN 			  & Datalink  & VLAN identity and priority \\
\midrule[0.3pt]
	MPLS			& Network & MPLS label and traffic class \\
\midrule[0.3pt]
	IP				& Network & IPv4 / IPv6 addresses \\
\midrule[0.3pt]
	Transport		& Transport & TCP/UPD, source and destination port \\

\bottomrule[1.2pt] \\
\end{tabular}
\end{table}

From OpenFlow protocol version 1.1 and onwards, Group Tables have
been defined. Group Tables allow more advanced configurations and
consist of three parameters:
\begin{itemize}
\item \emph{Action Buckets} - Each bucket is coupled to a switch port and
contains a set of Actions to execute. The main difference with Instructions
from the Flow Table is that Action Buckets can be coupled to counters
and interface status flags. Based on values of these parameters a
bucket is valid or not;
\item \emph{Type} - Defines the behavior and the number of Action Buckets
in the Group Table. Multiple Action Buckets can be used for i) multicast
and broadcast applications, where the incoming packet is copied over
multiple Action Buckets (multiple ports), ii) load sharing applications,
where a selection mechanism selects the Action Bucket to execute and
iii) failover applications, where from the available Action Buckets
the first live one is selected to execute. Assigning a single Action
Bucket to a Group Table is useful for defining Actions for a large
number of Flow Entries with the same required forwarding policy;
\item \emph{Counter} - The number of times the Group Table has been addressed.
\end{itemize}
\begin{figure}
\includegraphics[width=1\columnwidth]{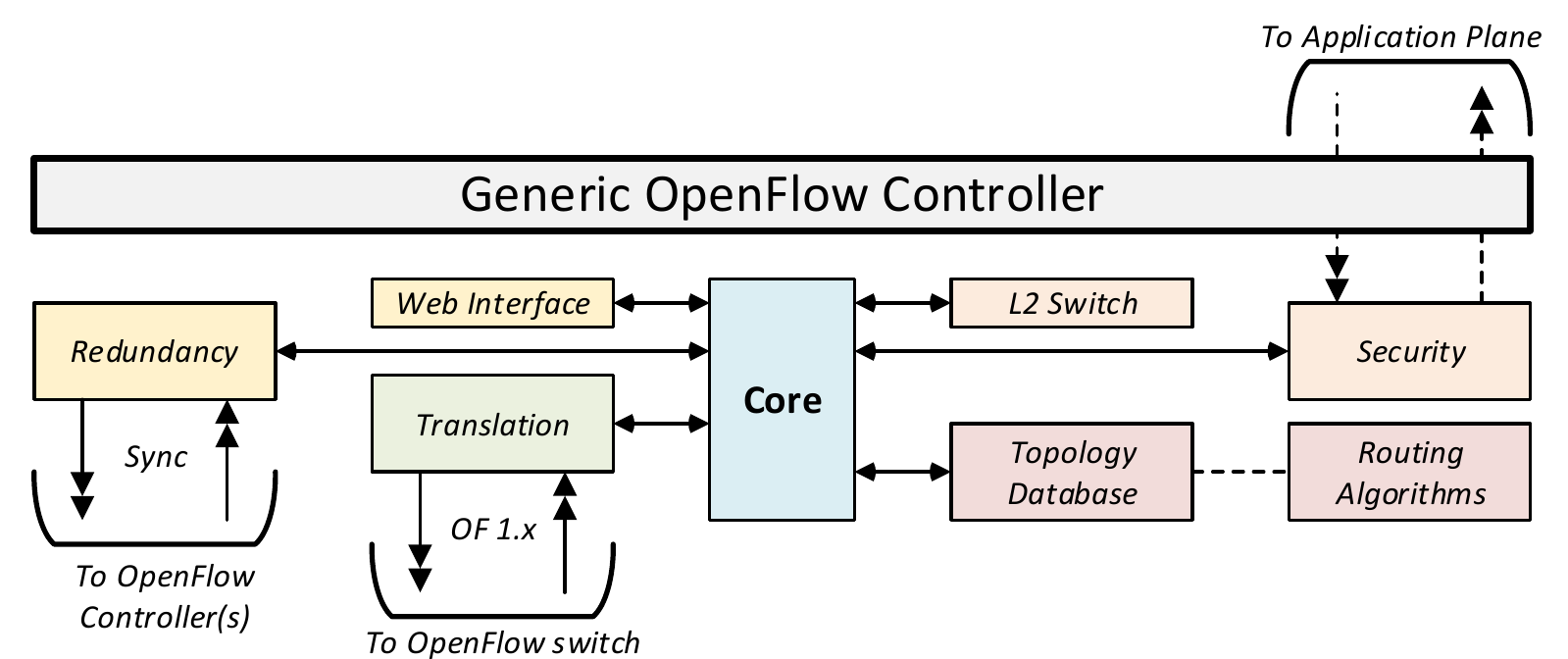}\protect\caption[Generic description of an OpenFlow Controller]{Generic description of an OpenFlow Controller - \emph{The controllers
mentioned are built around a Core application, which acts as a backbone
in the controller. To communicate with OpenFlow switches, a translation
module translates the OpenFlow protocol parameters to the ``language''
used inside the controller. Additional modules can advertise themselves
to the Core application to receive switch events. Based on the application,
flow rules can be calculated or notification are sent to the application
plane.} }

\label{fig:OpenFlow-Controller}
\end{figure}

With the descriptions of the Flow and Group Table we can follow the
incoming packet from figure \ref{fig:OpenFlow-Description}. At entry,
a metadata set is assigned to the data packet and the packet header
is matched to the Flow Entries in the first Flow Table. On match,
instructions are added to the Action Set and the packet can be processed
further. When the instructions include forwarding to other Flow Tables,
the packet with metadata is processed in a similar way and instructions
are added to the Set. When no forwarding to other Flow Tables is instructed,
the Action Set from the metadata is executed. Actions from the Set
and / or Group Table determine the process of the packet. In switching
operation, the MAC address is matched in the first Flow Table and
the Action Set defines how to forward the packet on a specified outgoing
port. When none of the Flow Entries match, a Table Miss is initiated.
Depending on the configuration by the OpenFlow controller, the packet
is dropped or transmitted to the controller for a Flow Request. At
the controller, new Flow Entries are computed and added to Flow Tables
of involved switches.

\subsection{OpenFlow Controller}

\label{sub:Description-of-OpenFlowController}

OpenFlow controllers are developed in many variations and all share
the same goal of controlling and configuring compliant switches. In
\cite{mendoncca2013survey} and \cite{lara2013network} a list of
hardware switches is given, all OpenFlow compliant. Main differences
are found in programming languages and support for OpenFlow specifications.
Popular implementations, like NOX \cite{gude2008nox} and the Open
vSwitch (OVS)-controller from Open vSwitch \cite{OVS2013ConfdB} use
the C/C++ language, while POX \cite{POXController} and Ryu \cite{RyuController}
are Python-based controllers. Java based controllers are found in
FloodLight \cite{FloodController} and OpenDayLight \cite{ODLController}.
Only Ryu, OpenDayLight and unofficial ported versions from NOX support
OpenFlow protocol version 1.3 so far. For more advanced configuration
purposes and the use of Group Tables, we advise the Ryu and OpenDayLight
controller. Both FloodLight and OpenDayLight offer web browser based
configuration tools instead of command line interfaces and are therefore
more user friendly. NOX, POX and Ryu share a similar structure and
this shared structure is used to give an example of an OpenFlow controller
in figure \ref{fig:OpenFlow-Controller}.

\begin{figure}[t]
\includegraphics[width=1\columnwidth]{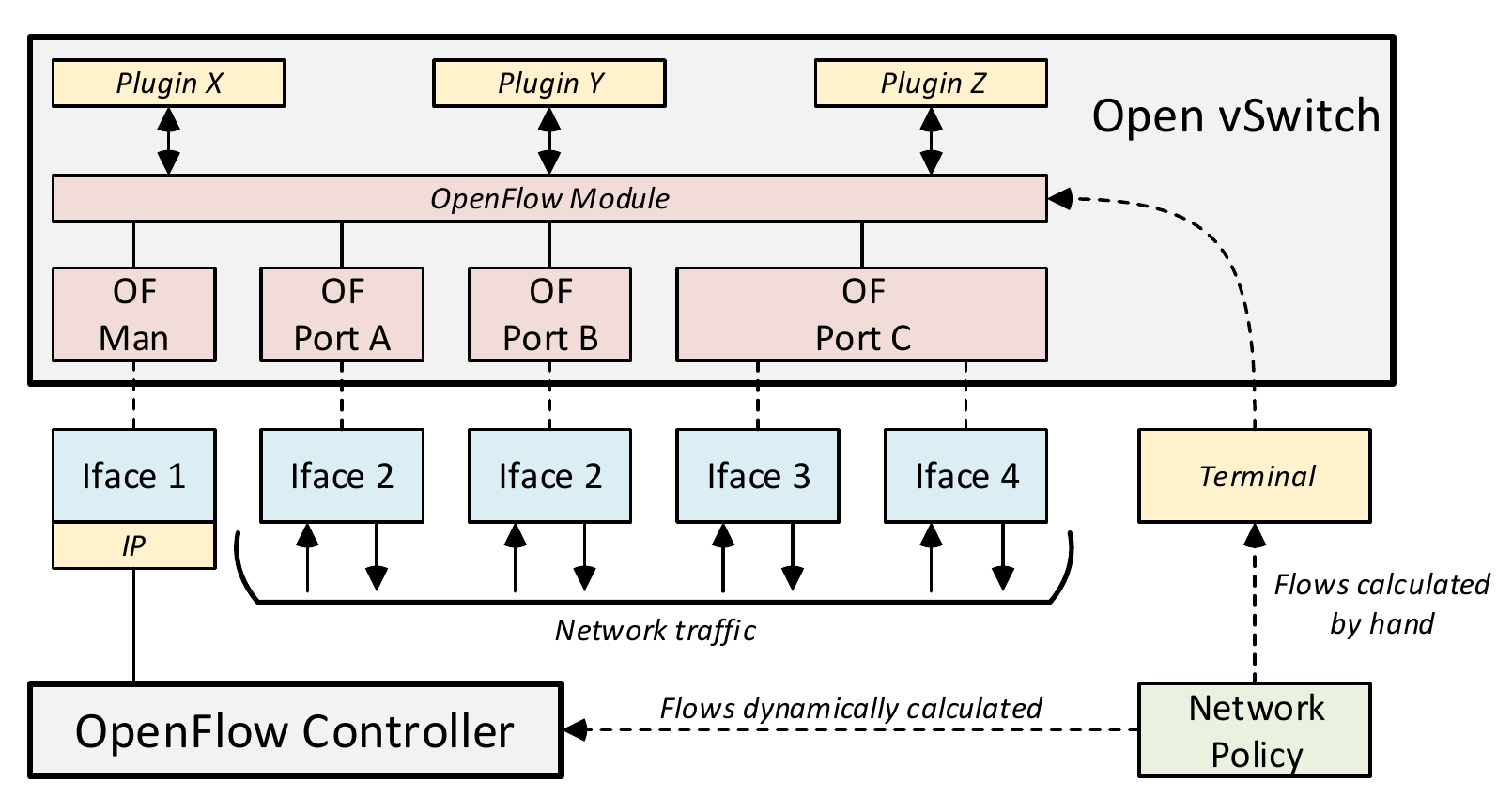}\protect\caption[Description of Open vSwitch in relation to OpenFlow]{Description of Open vSwitch in relation to OpenFlow - \emph{With
Open vSwitch physical network interfaces from servers and computer
can be assigned to an OpenFlow switch. To connect the virtual OpenFlow
switch to a remote controller via an IP connection, a management interface
(OF Man) with IP address is assigned. The other option to configure
the OpenFlow module is via a command line interface. Other physical
interfaces can be assigned directly to an OpenFlow port, where it
is also possible to bind multiple physical interfaces to a single
OpenFlow port. With other Open vSwitch plug-ins it is possible to
control the OpenFlow module.}}

\label{fig:OVS-Description}
\end{figure}

The example controller is built around a Core application, which acts
as an backbone in the controller. To communicate with the OpenFlow
switch, a translation module is added to translate OpenFlow protocol
messages to controller parameters. Other modules, like a layer-2 switch
module (L2-switch) in Ryu, can advertise themselves to the Core application
and register on specific switch parameters and events. The mentioned
controllers supply the Core application with translation modules for
OpenFlow protocol version 1.x. Depending on the requirements on the
controllers, one can construct and program modules and advertise these
to the controller. In figure \ref{fig:OpenFlow-Controller} examples
are given, such as a topology module, for maintaining an up-to-date
status of the network infrastructure. Also modules for redundancy
purposes can be added, to synchronize information with other controllers
at the control plane.

\subsection{Open vSwitch}

\label{sub:Description-of-OVS}

Although Open vSwitch (OVS) is not specifically designed to enable
the SDN philosophy, it is widely used by researchers and organizations
to test OpenFlow implementations and benefit from flexible SDN configurations.
OVS can be configured to turn regular servers and computers with multiple
physical network interfaces into a virtual OpenFlow switch, as shown
in figure \ref{fig:OVS-Description}. Many Linux distributions, such
as the Ubuntu OS, support OVS installation from their repositories.

Depending on the required configuration, OVS can be configured as
a layer-2 switch (controlled by a local OVS-controller) or as a generic
OpenFlow switch. Configuration of the OpenFlow module can be supplied
by an external OpenFlow controller or Flow Rules are supplied manually
via the command line interface. External plug-ins can also configure
the OpenFlow module of OVS. An example of such a plug-in is the Quantum
plug-in from OpenStack \cite{OpenStack}.

\section{Graphical SDN Framework}

\label{sec:OverviewFramework}

\begin{figure}
\includegraphics[width=1\columnwidth]{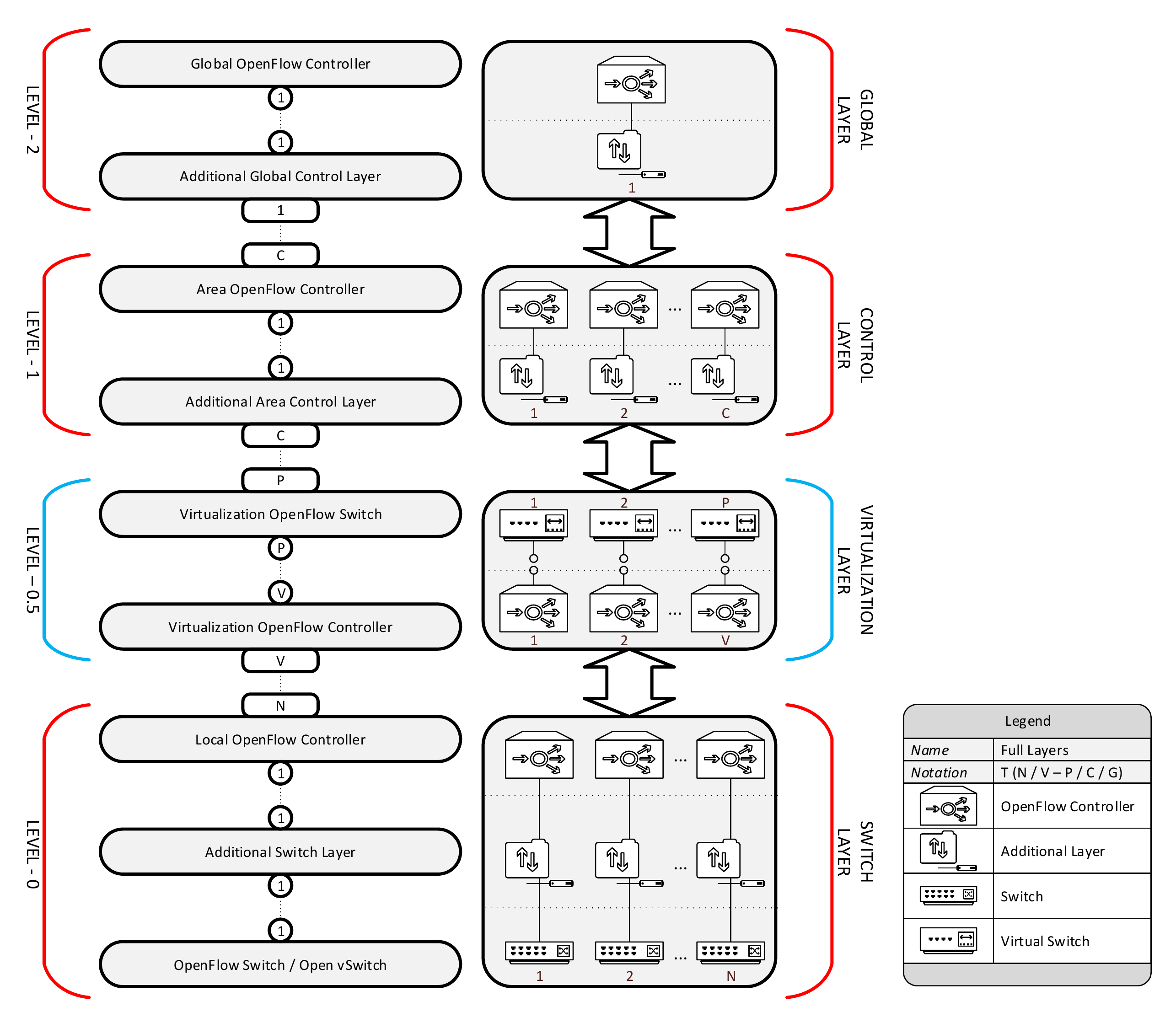}\protect\caption[Graphical Framework for OpenFlow differentiation and comparison]{ Graphical Framework for OpenFlow differentiation and comparison
- \emph{On the left the numerical UML relationship between the components
and layers of an OpenFlow network topology is visible. On the right
a physical decomposition of the same configuration is given to clarify
the UML relationship.}}

\label{fig:Graphical-Framework}
\end{figure}

To differentiate and compare the existing SDN solutions, we have developed
a graphical SDN framework. Within the graphical framework multiple
layers are defined, which indicate the hierarchy level of components
within SDN networks (see figure \ref{fig:Graphical-Framework}).

For our graphical framework we define that controllers perform the
computations and tasks to control traffic and additional layers can
be added for administrative and synchronizing purposes. In the following
we explain the layers of figure \ref{fig:Graphical-Framework}.
\begin{itemize}
\item Level-0 - \emph{Switch Layer} - The lowest layer identified in the
OpenFlow structure is the switch layer, with the main purpose to deliver
data plane functionality. Data plane functions are performed at the
\emph{Switch / Open vSwitch} sublayer, where the two additional sub-layers,
being the \emph{Additional Switch Layer} and\emph{ Local OpenFlow
Controller}, add additional functionality to perform minor control
plane tasks;
\item Level-0.5 - \emph{Virtualization Layer} - On top of the switch layer,
the \emph{Virtualization Layer} can be placed with the main function
to divide and share the switch resources over multiple OpenFlow controllers.
It enables multiple virtual network topologies on top of a single
physical infrastructure. Resources of physical switches are virtualized
by this layer and presented to the \emph{Control Layer} as multiple
virtual switches; 
\item Level-1 - \emph{Control Layer - }The functionality of the control
layer is to perform the tasks of the SDN control plane in a defined
area of the network topology for a number of switches. Decisions made
at this layer influence only a part of the network and are locally
optimal. In regular OpenFlow configurations, only a single \emph{Area
OpenFlow Controller} is present. Solutions have been proposed to enhance
the control layer with additional area OpenFlow layers\emph{ }to extend
functionality, such as synchronization of Flow Rules with other controllers;
\item Level-2 - \emph{Global Layer} - The top layer has the functionality
to control network topology at a global level, where forwarding and
routing decisions influence the whole topology. A \emph{Global OpenFlow
Controller} can thus compute globally optimal routes through the network,
as it controls all switches. The structure of the global layer is
similar to the control layer.
\end{itemize}
To indicate the numerical relationship between the layers, the UML-standard,
as used in object-oriented programming, is used as guidance in the
framework in figure \ref{fig:Graphical-Framework}. The relationship
states that the components (sub-layers) at each level share a one-to-one
relationship ($1..1$) with each other. From the switch level a many-to-many
or many-to-few relationship exists with the virtualization layer ($N..V$)
or control layer ($N..C$), when no virtualization is applied. In
the case of virtualization, a many-to-few or many-to-many relation
($P..C$) indicates $P$ virtual switches are controlled by $C$ OpenFlow
controllers. Within a domain, multiple area controllers can be controlled
by a single centralized controller with the global view of the network.
In case of an inter-domain network infrastructure, global layers can
be interconnected.

\begin{table}
\protect\caption{Explanation for the developed OpenFlow notation standard}
\label{tab:StandardNotation}

\centering
\begin{tabular}{c l l}

	\emph{Symbol} & \emph{Description} & \emph{Relation} \\
\toprule[2pt]

	$N$ & No. Switches & $\in(1,2,..,N)$ 										  				\\

	$V$ & No. Virtual controllers & $\in(1,2,..,V)$,$V \leq N$					 				\\

	$P$ & No. Virtual switches & $\in(1,2,..P)$ 								   				\\

	$C$ & No. Area OpenFlow Controllers & $\in(1,2,..,C)$,$C \leq N$, 	 		\\

		&						& $C \leq V$ or $C \leq P$ 					\\

	$G$ & Global controller enabled & $\in(0,1)$ 							   				\\

\midrule[1pt]

	$X^{+s}$ & Layer enhanced for security &		\\
	$X^{+p}$ & Layer enhanced for scalability &	 \\
	$X^{+r}$ & Layer enhanced for resiliency &	  \\
	$X^{+b}$ & Backup component available &		 \\

\bottomrule[1.2pt] \\
\end{tabular}
\end{table}

In order to differentiate multiple network topologies using the UML
relationships, we use the following notation. For network topology
$T$ the notation is given as $T(N/V-P/C/G)$, where the description
of the used symbols is given in table \ref{tab:StandardNotation}.
The notation with the defined symbols of table \ref{tab:StandardNotation}
would not cover the entire framework, as additional sub-layers or
applications are not indicated. Therefore an extra indicator is added
to the OpenFlow notation, to indicate if an enhancement is added and
for which enhancement area (security, scalability and/or resiliency)
the enhancement is added. Beside additional components to the layers,
it is possible from OpenFlow protocol version 1.2 to add redundant
controllers to the control plane. Therefore the backup indicator is
defined. When an enhancement overlaps multiple areas or when a component
is applied redundantly, it is possible to combine indicators. The
controller $C^{+sb}$ indicates a security enhanced controller that
is redundantly applied.

\begin{figure}
\includegraphics[width=1\columnwidth]{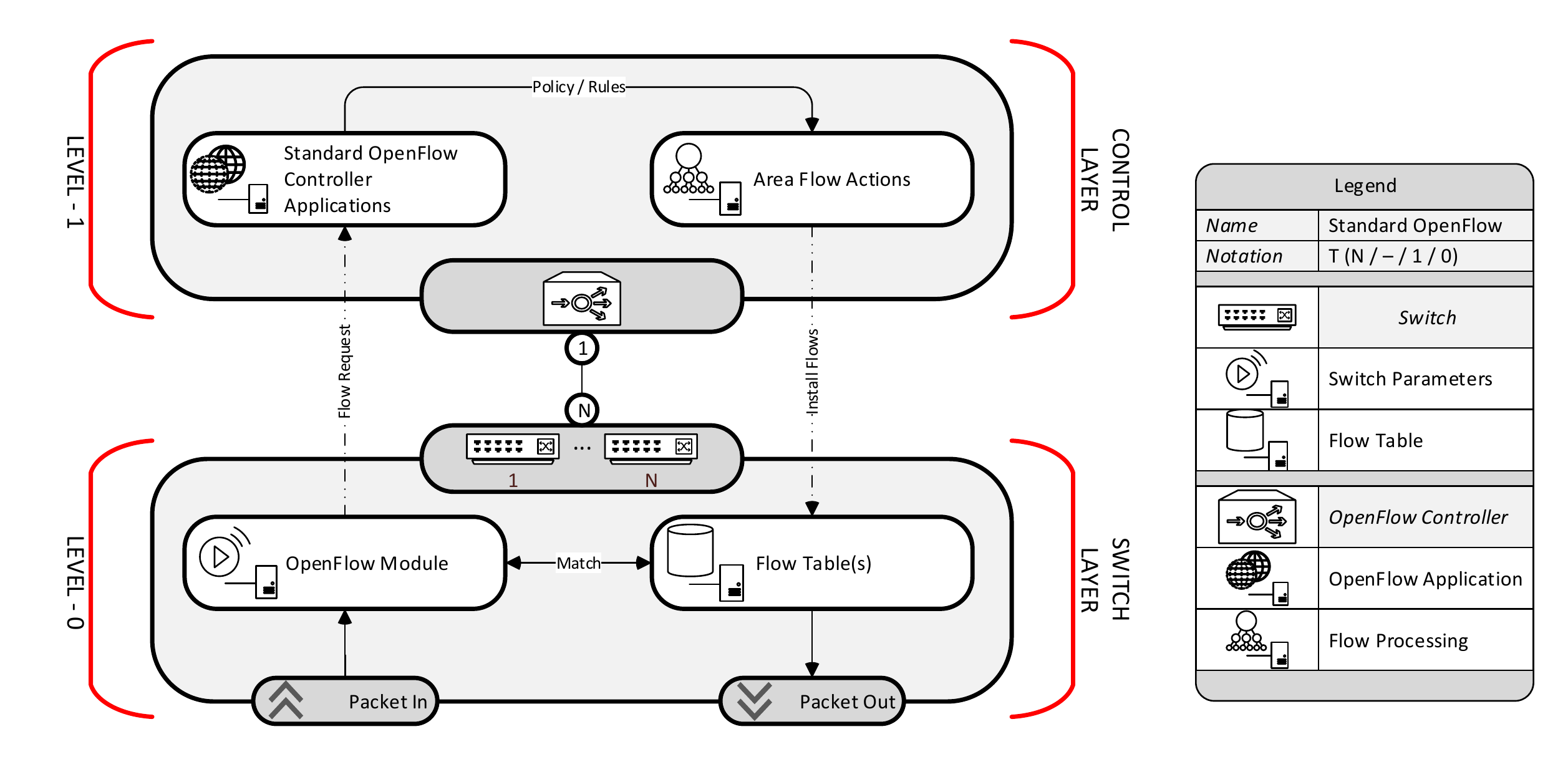}\protect\caption[Reference OpenFlow configuration]{Reference OpenFlow configuration - \emph{Arriving data packets are
processed and on a Table Miss, a Flow Request is sent to a }``\emph{standard''
OpenFlow controller, where network routing applications determine
a new Flow Rule based on set Policy and Rules. The new Flow Rule is
installed by the controller in the assigned switches and network traffic
can transverse the network.}}

\label{fig:Framework-Standard}
\end{figure}

Before multiple OpenFlow enhancement proposals will be discussed,
a reference OpenFlow network configuration is given in the figure
\ref{fig:Framework-Standard}. The reference configuration is indicated
by $T(N/-/C/0)$, where the infrastructure is built up from $N$ switches
controlled by $C$ general OpenFlow controllers.

The flow of actions in figure \ref{fig:Framework-Standard} is as
follows. Network traffic, in the form of data packets, arrives at
the switch data plane, where the packet headers are matched to a set
of flow rules stored in the flow tables. When no match is found, the
OpenFlow module of the switch will initiate a flow request to the
assigned controller. The controller will process the event and install
the needed flow rules into the switches at the designated route through
the network, based on the policy and rules defined in the forwarding
and routing applications of the controller.

\section{Scalability in SDN}

\label{sec:Scalability-in-SDN}

By introducing a centralized authority to control the network traffic
over a large network, the growth of network traffic may not scale
with the performance of the controller \cite{yeganeh2013scalability}.
Multiple proposals have been introduced \cite{curtis2011devoflow,tootoonchian2010hyperflow,hassas2012kandoo,koponen2010onix,sherwood2009flowvisor}
to create a distributed centralized authority, to solve the concerns
on scalability and performance. In this section, these proposals will
be discussed and projected onto our developed graphical framework.

\subsection{HyperFlow}

\label{sub:HyperFlow}

The proposed solution by Tootoonchian et al. \cite{tootoonchian2010hyperflow}
to solve the scalability problem in SDN and OpenFlow networks is to
deploy multiple OpenFlow controllers in the network infrastructure
and implement a distributed event-based control plane, resulting in
a $T(N/-/C^{+p}/1)$ structure. The improvement in scalability and
performance is found in the placement of multiple controllers close
to the switches, reducing flow setup times, while each controller
is provided with the global view of the network. Another advantage
of this approach is the resiliency against network partitioning (disconnected
network) in case of network failures, as the views of the connected
switches and controllers are synchronized. 

\begin{figure}
\includegraphics[width=1\columnwidth]{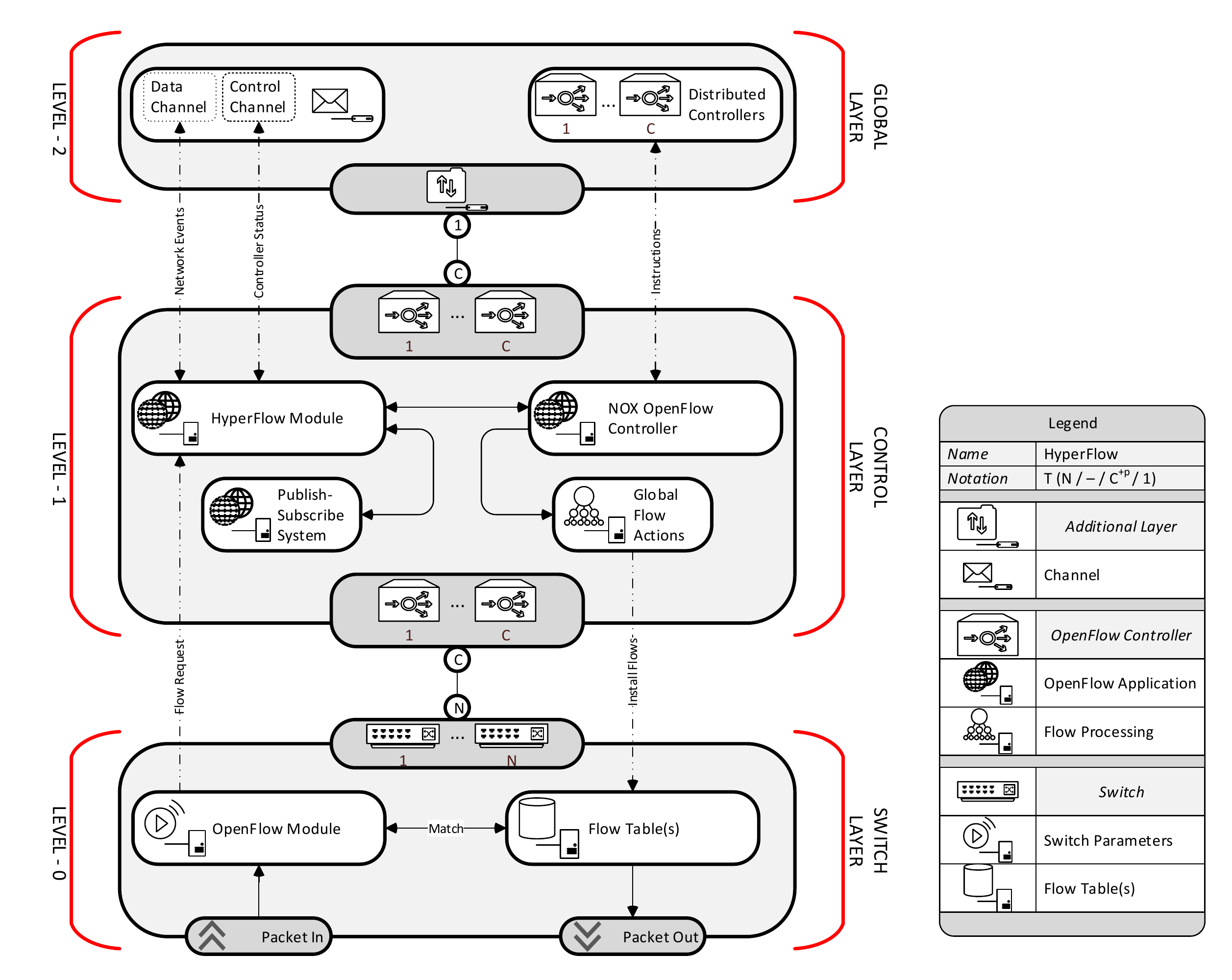}\protect\caption[HyperFlow implementation]{HyperFlow implementation - \emph{An additional global layer and extra
OpenFlow applications at the control layer provide area controllers
with additional information, to increase decision making capabilities
at the global level for optimal data packet forwarding.}}

\label{fig:Framework-HyperFlow}
\end{figure}

HyperFlow, see figure \ref{fig:Framework-HyperFlow}, is built up
from three layers, where no virtualization is applied. The architecture
at the switch level (data plane) and the OpenFlow specification are
unmodified, which makes the HyperFlow concept applicable to current
SDN-capable networks. At the control layer the HyperFlow application
is connected to a standard NOX controller. To enable communication
with the global layer, a publish-subscribe system is placed at the
additional control layer to locally store the network status and view.
The HyperFlow distributed global layer consists of three parts: the
data channel, control channel and the distribution of functionality
among the controllers. On the data channel, network events are distributed
by the HyperFlow application. These events are stored in the publish-subscribe
system, which synchronizes the global view of the controllers, but
can operate independently in case of partitioning. The control channel
is used for monitoring the status of the controllers. The distribution
of functionality among the controllers is realized as follows. Each
controller has a global view of the network and its status, it has
capabilities to install flows into all switches. All required flows
for a particular data stream are published to all distributed controllers
to synchronize network state. The HyperFlow application on the controllers
filter the requested flows for the switches assigned to it and will
install the flows accordingly.

In \cite{tootoonchian2010hyperflow} no extensive measurements are
performed on different network topologies and traffic loads, thus
no conclusion can be drawn from the HyperFlow approach to improve
scalability in real-life SDN networks. Besides that, there are some
limitations in the HyperFlow design. The first limitation was found
by the authors themselves in the performance of the publish-subscribe
system (WheelFS). All network events, flow installs and status information
need to be synchronized between the multiple controllers, which requires
a fast distributed storage system. In HyperFlow the performance of
the publish-subscribe system was limited to approximately $1000$
events per second. This does not indicate that the controllers are
limited in processing, but the global view of the network controllers
may not quickly converge. A second limitation is the lack of a management
application in the global layer. In \cite{tootoonchian2010hyperflow}
no distinction is made between the functionality of the switches and
controllers. This assumes that a global policy for Flow Rule installations
must be configured in all assigned controllers. The last limitation
is found in the performance of HyperFlow, where network traffic may
be dropped when the offered load exceeds a controller's capacity.
Although the load on controllers can be reduced by assigning less
switches to a controller, a single switch processing many flows may
still overload its controller. We think a solution where Flow Requests
can be forwarded to neighbor controllers for processing or smart applications
at the switch layer to off-load controllers could be another solution.

\subsection{ONIX}

\label{sub:ONIX}

\begin{figure}
\includegraphics[width=1\columnwidth]{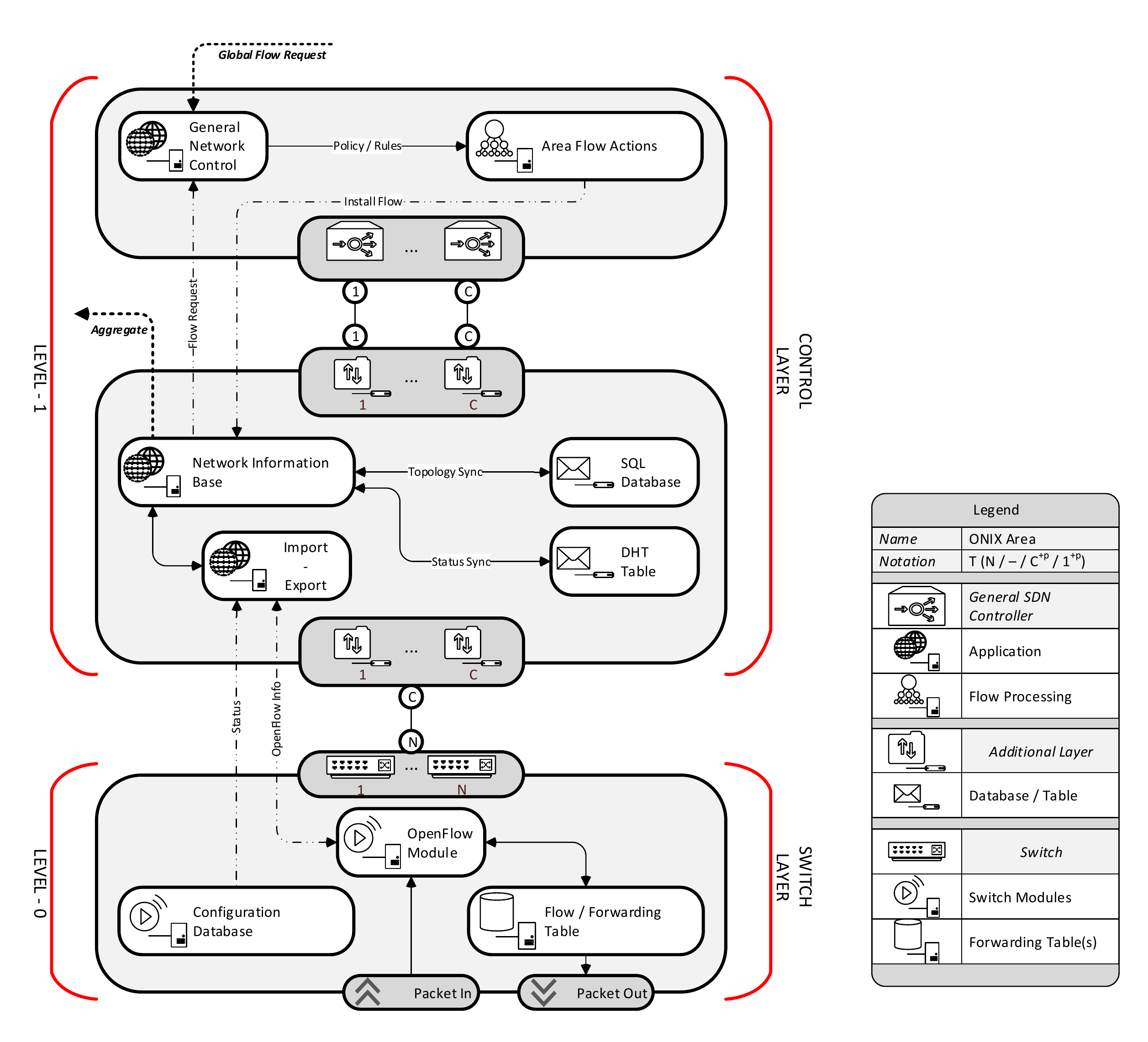}\protect\caption[ONIX distributed control platform at area level]{ONIX distributed control platform at area level - \emph{ONIX adds
a distributed control platform between the data and control plane
to reduce the workload on flow controllers. Switch management is performed
by multiple ONIX instances which are synchronized via two databases.
Forwarding decisions are made at area level by a general SDN flow
controller.}}
\label{fig:ONIX-Area}
\end{figure}

\begin{figure}
\includegraphics[width=1\columnwidth]{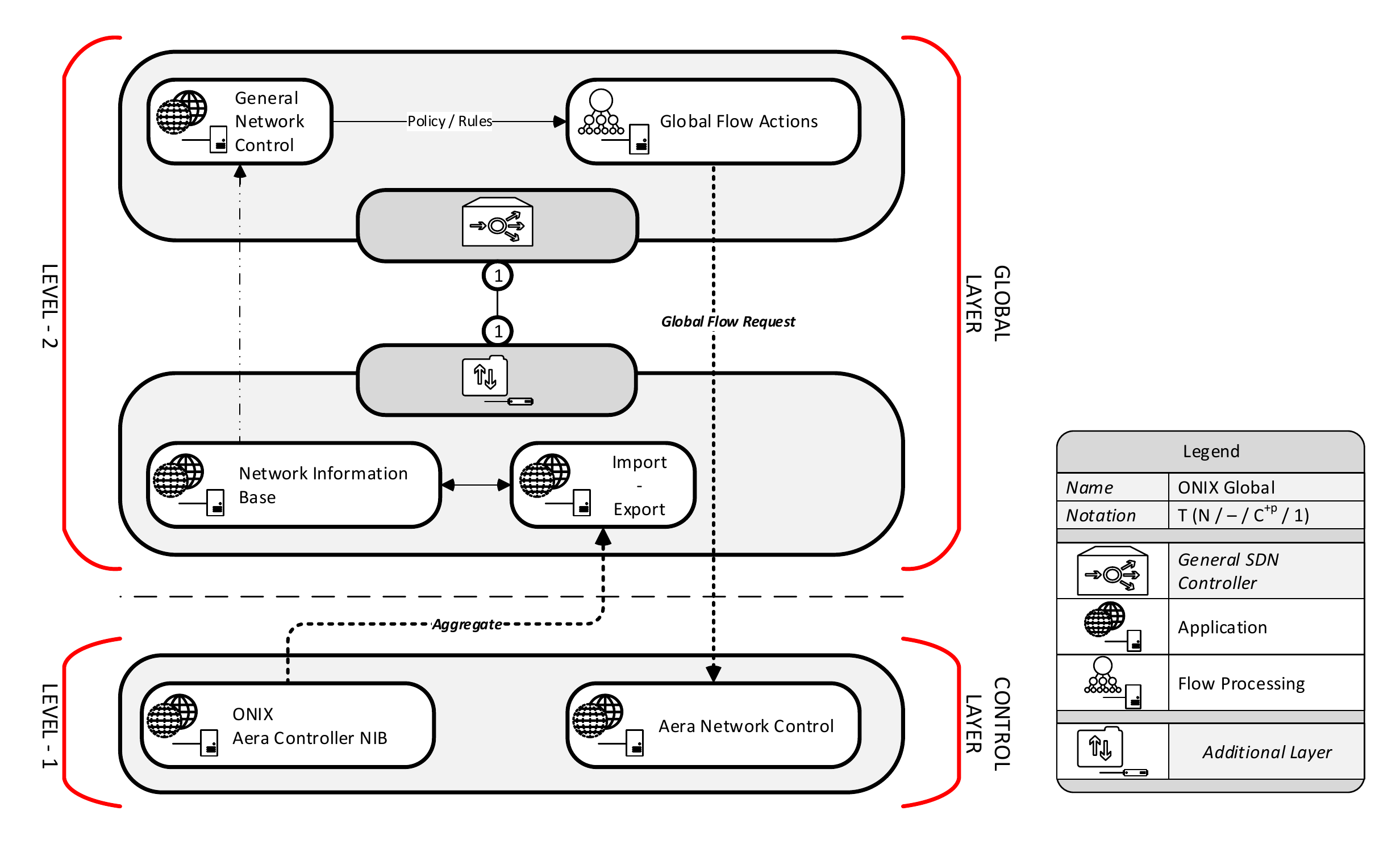}\protect\caption[ONIX distributed control platform at global level]{ONIX distributed control platform at global level - \emph{Aggregate
information from multiple area controllers are combined at the ONIX
global layer, from where the global controller can determine optimal
routing and request the level-1 controllers for paths within the assigned
area.}}
\label{fig:ONIX-Global}
\end{figure}

ONIX \cite{koponen2010onix} is not built on-top of an OpenFlow controller,
but can be classified as a ``General SDN Controller.'' The approach
the developers took to abstract the network view and overcome scalability
problems, was to add an additional distributed control platform. By
adding an additional platform, the management and network control
functions from the control plane are separated from each other. The
management functions, used for link status and topology monitoring,
are performed by the ONIX control platform and reduce the workload
for a general flow controller. To partition the workload at the control
platform, multiple ONIX instances can be installed to manage a subset
of switches in the network infrastructure. The available information
is synchronized between the ONIX instances, forming a distributed
layer as seen with HyperFlow in section \ref{sub:HyperFlow}, but
now at the controller layer. A network controller assigned to a part
of the network, can subtract information from the distributed layer
to calculate area flows. In order to calculate global flows, ONIX
has the capability to aggregate information from multiple area platforms
to a global distributed platform. The aggregation of information and
the calculation of global optimal forwarding rules is similar to the
routing of internet traffic over multiple autonomous systems (ASes)
\cite{rfc1930}, where the route between ASes is globally determined,
but the optimal route inside the autonomous system is calculated locally.
With this general introduction, we can conclude that ONIX can be classified
as a $T(N/-/C^{+p}/1^{+p})$ SDN topology%
\footnote{In this overview the network scope is limited to a single domain,
but the capabilities of ONIX can reach beyond that scope as it is
designed for large-scale production networks.%
}. With the use of figure \ref{fig:ONIX-Area} more insight is given
in the design and functioning of ONIX at area level, while figure
\ref{fig:ONIX-Global} shows how global flows are calculated using
the distributed control platform.

At switch layer no modifications are required for ONIX and two channels
connect to the switch with the Import-Export module at the distributed
platform. The ONIX platform is an application that runs on dedicated
servers and requires no specific hardware to function. According to
the load and traffic intensity, multiple switches are assigned to
an ONIX instance. Multiple ONIX instances combined form the distributed
control platform. The first channel connects the configuration database
and is used for managing and accessing general switch configuration
and status information. To manage the forwarding, flow tables and
switch port status, the second channel is connected to the OpenFlow
module of the switch. The information and configuration status collected
by the import-export module represent the network state of the connected
switches and is stored in the Network Information Base (NIB) as a
network graph. The NIB uses two data stores, being a SQL-database
for slow changing topology information and a Dynamic Hash Table (DHT-table)
for rapid and frequent status changes (such as link utilization and
round trip times). The application of two data stores overcomes the
performance issues faced by HyperFlow. From the NIB, a general flow
controller or control logic can receive flow requests, switch status
and network topology and compute paths. In comparison to the reference
OpenFlow controller (figure \ref{fig:Graphical-Framework}), the ONIX
platform has the availability of switch and link status information.
Computation of paths is thus not limited by the information provided
by the OpenFlow protocol. The last step in the flow setup process
is to install the computed flows in the switch forwarding table via
the ONIX distributed platform.

In \cite{koponen2010onix} evaluation results are shown of measurements
on the ONIX distributed layer to determine the performance of a single
ONIX instance, multiple instances, replication times between the data
stores and recovery times on switch failures. Unfortunately no information
is present of the used control logic and the performance gain in comparison
with regular OpenFlow controllers. The advantage of the ONIX distributed
control platform is the partitioning of the workload over multiple
instances. So, if an ONIX instance is limiting traffic throughput
(dropping flow requests) due to high workload, assigned switches can
be reassigned to other ONIX instances.

\subsection{DevoFlow}

\label{sub:DevoFlow}

The approach of DevoFlow \cite{curtis2011devoflow} is guided by two
observations. First, the amount of traffic between the data and control
planes needs to be reduced, because the current hardware OpenFlow
switches are not optimized for inter-plane communication. Second,
the high number of flow requests must be limited, because the processing
power of a single OpenFlow controller may not scale with network traffic.
To limit the traffic and flow requests to the control plane, traffic
may be categorized into \emph{micro} and \emph{elephant} flows. In
DevoFlow micro flows will be processed at the switch layer, without
the need of the control layer. The DevoFlow philosophy is that only
heavy traffic users, elephant flows, need flow management. By limiting
the management to elephant flows, only one controller is needed in
the network topology, shifting the routing complexity to the switch
layer. With this information we classified the DevoFlow solution as
a $T(N^{+p}/-/1/0)$ SDN topology.

\begin{figure}
\includegraphics[width=1\columnwidth]{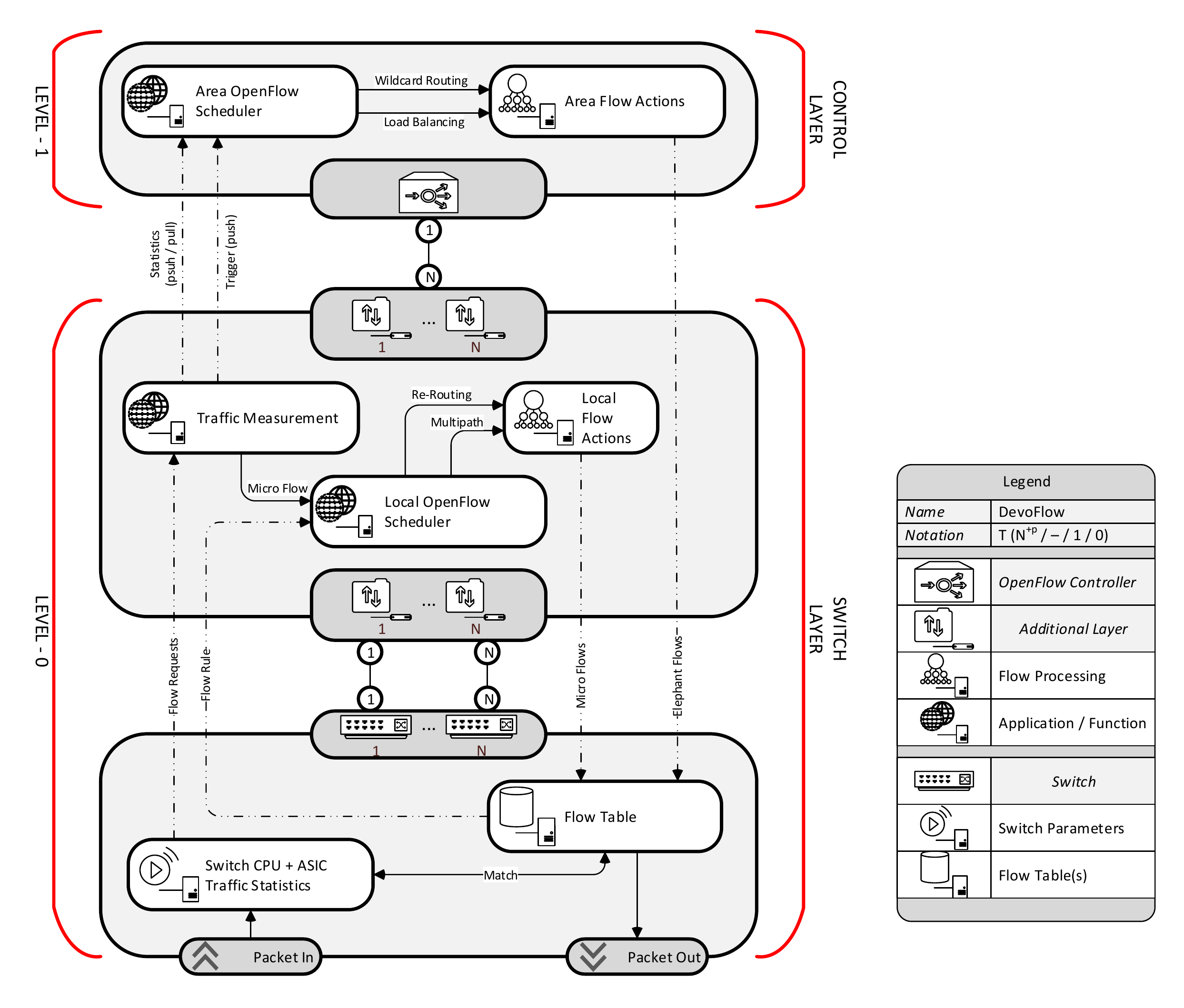}\protect\caption[DevoFlow implementation]{DevoFlow implementation - \emph{Scalability enhancements made at
the switch layer are shown as an additional layer, but are implemented
as modifications to the soft- and firmware in the switch ASIC / CPU.
Routing decisions can be made at the switch layer using traffic parameters
as input to reduce workload at the OpenFlow controller. Elephant flows
invoke the use of the area controller for optimal routing, while micro
flows are routed using cloned flow rules.} }
\label{fig:DevoflowFramework}
\end{figure}

In figure \ref{fig:DevoflowFramework}, the DevoFlow solution is drawn
as an additional layer on top of the physical switch to simplify the
representation, but the actual implementation is performed at the
soft- and firmware of the switch. This indicates that modifications
to the standard switch are required. To ease integration of DevoFlow
with other SDN concepts, no modifications are made to the OpenFlow
protocol and controller. The destination of arriving packets is compared
to flow rules installed in the switch table. When no match is found,
a flow request must be initiated by the ``Traffic Measurement'' module
in the switch. The traffic measurement module monitors the packets
in the data flows and computes statistics. At the start of a flow
each flow is marked as a micro flow and an existing flow rule from
the forwarding table is cloned and modified by the ``Local Flow Scheduler.''
The modification to the flow rules allows multipath routing and re-routing.
In case that multiple paths between the switch and destination exist,
the local flow scheduler can select one of the possible ports and
the micro-flow rule for that port is cloned. Re-routing is applied
when one of the available switch ports is down and traffic needs alternative
paths through the network.

To detect and route elephant flows through the network, the area scheduler
can use four different schemes:
\begin{itemize}
\item \emph{Wild-card routing} - The switch pushes the available traffic
statistics to the controller with a specified time interval. The scheduler
pro-actively calculates unique spanning trees for all destinations
in the network topology using the least-congested route and install
the trees as flows in the switch flow tables. So for each destination
in the network a flow is present in the switches and no flow requests
are needed from the switch to the OpenFlow controller;
\item \emph{Pull-based statistics - }The scheduler regularly pulls the traffic
statistics from the switch and determines if elephant flows are present
in the current data flows. Once an elephant flow is detected, the
scheduler determines the least-congested path for this flow and installs
the required flow rules at the switches;
\item \emph{Sampled statistics - }This method is very similar to the pull-based
scheme, but instead of pulling traffic statistics every time period,
the switch samples traffic statistics into a bundle and pushes the
bundle to the scheduler. At the scheduler, it is again determined
if any elephant flows are present and on positive identification flows
are installed, as described in the pull-based scheme;
\item \emph{Threshold - }For each flow at the switch, the amount of transferred
data is monitored. Once a flow exceeds a specified threshold, a trigger
is sent to the scheduler and the least-congested path is installed
into the switches.
\end{itemize}
All schemes are based on traffic statistics, where flows are only
installed if identified as elephant flows in a reactive manner. In
\cite{curtis2011devoflow} multiple simulations have been performed
on a large data network simulator to capture the behavior of flows
through the network and measure data throughput, control traffic and
the size of flow tables in the switches. The results show that the
pull-based scheme with a short update interval maximizes the data
throughput in the network. This performance comes at a price, as much
traffic is initialized between the switch and controller and the size
of the flow table is significantly large in comparison with the other
schemes. The threshold scheme is identified as most optimal, as the
data throughput is high, less traffic is required between the switch
and the controller and the size of the flow table is minimal. Another
advantage is the required workload on the scheduler in the controller,
as no traffic statistics have to be monitored and processed.

\subsection{Kandoo}

\label{sub:Kandoo}

\begin{figure}
\includegraphics[width=1\columnwidth]{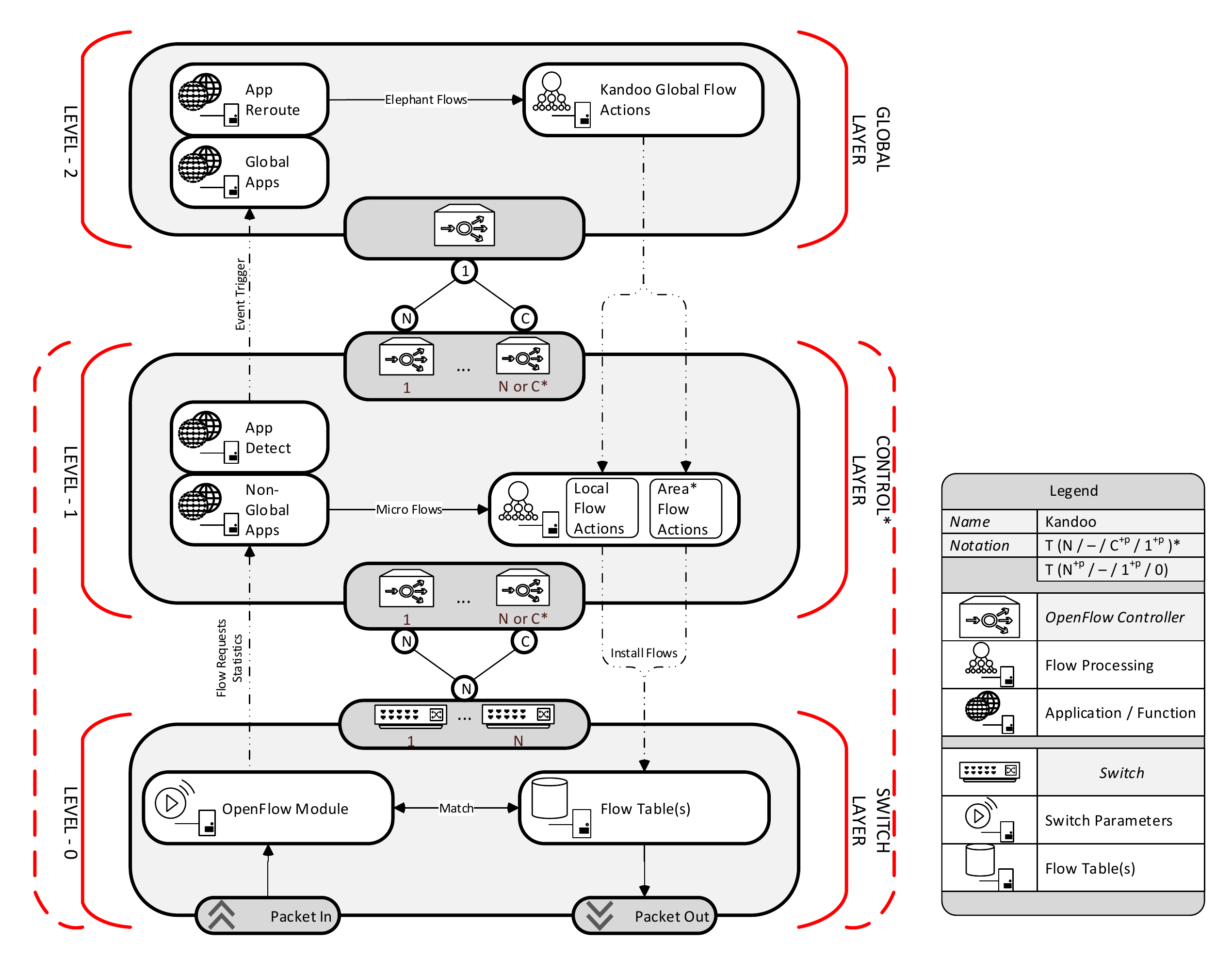}\protect\caption[Kandoo implementation]{Kandoo implementation - \emph{A standard switch sends flow requests
and statistics to the local controllers. The Non-global applications
process local OpenFlow events, micro flows and the ``App Detect''
application detects elephants flows using a threshold. Elephant flows
are processed by the root / global controller, where after the flow
rules are sent to the local controller to install at the switches.}}
\label{fig:KandooFramework}
\end{figure}

In \cite{hassas2012kandoo} Kandoo is presented with similar goals
and philosophy as DevoFlow, namely to limit the overhead of events
between the data and control planes, but solves the problem by applying
more controllers in a network topology. Kandoo differentiates two
layers of controllers, namely local controllers, which are located
close to the switches for local event processing, and a root controller
for network-wide routing solutions. In practice this means that local
controllers will process the micro flows and a trigger from the local
controller must inform the root controller about the presence of an
elephant flow. The number of local controllers depends on the amount
of traffic (local events) and the workload on the controller, which
is somewhat similar to the approach of ONIX and its distributed control
platform. In an extreme case, every switch is assigned to one local
controller. Translating Kandoo to the graphical framework, the root
controller is located at the global layer and the local controllers
can be found on the area or switch layer. If the extreme case is valid,
the local controller can be seen as an extension of the switch layer
and the topology can be classified as $T(N^{+p}/-/1/0)$. In a regular
architecture, where more than one switch is assigned to a local controller,
a $T(N/-/C^{+p}/1)$ SDN topology is found. Figure \ref{fig:KandooFramework}
projects the regular case of Kandoo.

Kandoo leaves the software in the switch unmodified and shifts the
processing of events to local controllers. The local controllers are
standard OpenFlow controllers extended with a Kandoo module. This
approach keeps the communication between the switch and controller
standardized and gives the possibility to utilize standard OpenFlow
applications. The Kandoo module intercepts the OpenFlow traffic and
monitors it for elephant flows using the ``App Detect'' application.
As long as no elephant flow is detected, the local controller processes
the flow requests as micro flows. Elephant flows are detected using
the threshold scheme from DevoFlow, which relays an event to the Kandoo
module of the root controller in case of positive detection.

To propagate events from the local controllers to the root controller,
a messaging channel is used. The root controller must subscribe to
this channel in order to receive events. After receiving an elephant
flow detection trigger, the ``App Re-Route'' determines an optimal
route and requests the local controllers to install this route. The
developers of \cite{hassas2012kandoo} have not given any information
on how the re-routing process is executed and which routing schemes
are used. 

Some measurements have been performed on a (small) tree topology with
the Kandoo framework installed. Results show comparisons between the
topology in a standard OpenFlow and Kandoo configuration. As expected,
less events are processed by the root controller, but no information
is given about the workload and performance of the local controllers.
Overall we can state that the simulations and measurements are too
limited to give a good indication of the performance enhancement provided
by Kandoo. The limiting factor in the Kandoo configuration is the
interface between the switch and the local controller, as in DevoFlow
is shown that this interface is the limiting factor in current SDN
network implementations. If the limit on this interface is not reached,
the layered controller solution is an interesting concept which can
also be useful to tackle security and resiliency problems.

\subsection{FlowVisor}

\label{sub:FlowVisor}

\begin{figure}
\includegraphics[width=1\columnwidth]{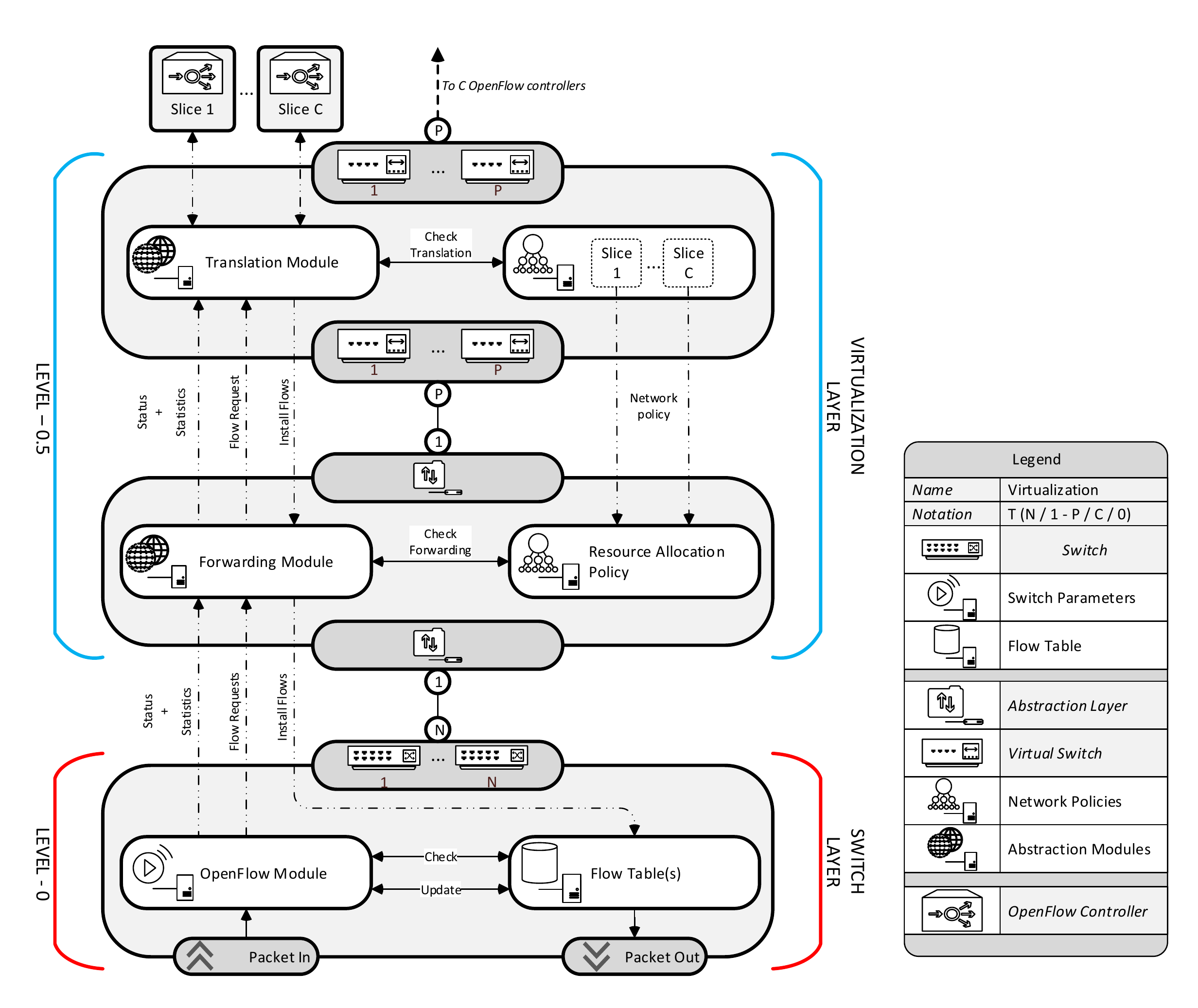}\protect\caption[Network virtualization on OpenFlow network topology]{Network virtualization on OpenFlow network topology - \emph{Regular
OpenFlow traffic (Flow Requests, status and traffic statistics) from
the OpenFlow module is sent to the FlowVisor Forwarding Module. Depending
on the slice configuration and slice policy, OpenFlow traffic is forwarded
to the Translation Module of the }``\emph{virtual'' switches. Guest
controllers (which can be any OpenFlow controller) communicate with
the translation modules, where Flow Rules are translated, forwarded
and installed if they do not interfere with the network policy for
that slice.}}
\label{fig:SDNR-FlowVisor}
\end{figure}

Virtualization is a widely applied technique allowing multiple instances
on the same hardware resources. Hardware resources are abstracted
by an abstraction layer and presented to a virtualization layer. On
top of the virtualization layer, instances are presented with virtualized
hardware resources that one can control as if without virtualization.
This approach is roughly similar to the SDN philosophy, but in FlowVisor
by Sherwood et al. \cite{sherwood2009flowvisor} the virtualization
approach is reapplied, where OpenFlow-compliant switches are offered
to the FlowVisor abstraction layer. FlowVisor offers \emph{slices}
of the network topology to multiple OpenFlow \emph{guest} controllers,
where the slices are presented as virtual OpenFlow switches. The guest
controllers control the slices, where FlowVisor translates the configurations
and network policies from each slice to Flow Rules on the physical
OpenFlow switches. If this approach is applied to large networks,
scalability problems can be resolved as control of the OpenFlow switches
is divided over multiple controllers. To distinguish network slices,
four dimensions are defined in FlowVisor:
\begin{itemize}
\item \emph{Slice} - Set of Flow Rules on a selection of switches of the
network topology to route traffic; 
\item \emph{Separation} - FlowVisor must ensure that guest controllers only
can control and observe the assigned part of the topology;
\item \emph{Sharing} - Bandwidth available on the topology must be shared
over the slices, where minimum data rates can be assigned for each
slice;
\item \emph{Partitioning} - FlowVisor must partition the Flow Tables from
the hardware OpenFlow switches and keep track of the flows of each
guest controller.
\end{itemize}
With these four dimensions and definitions, OpenFlow resources can
be virtualized and shared over multiple instances. This means that
a single hardware OpenFlow switch can be controlled by multiple guest
controllers. To provide more details on the functioning of network
virtualization in OpenFlow, FlowVisor is represented in figure \ref{fig:SDNR-FlowVisor}.

As illustrated in figure \ref{fig:SDNR-FlowVisor}, FlowVisor acts
like a transparent layer between hardware switches and the controllers.
Hardware OpenFlow switches are assigned to the virtualization layer,
where FlowVisor advertises itself as a controller. OpenFlow traffic
is transmitted to FlowVisor at the virtualization layer, where the
network capacity is sliced and divided over multiple users. The FlowVisor
Forwarding Module checks on policy violations, before traffic is sent
to the Translation module at the virtual OpenFlow switch. At the translation
module, the traffic is translated and sent to the guest controllers,
assigned to the corresponding slices. Flows coming from the guest
controllers are checked on interference with the slice policy and
translated to flows for the hardware switches. Via the forwarding
module the guest flows are installed at the switches.  The projection
shows a straight forward example of a $T(N/1-P/C/0)$ configuration,
where one FlowVisor instance performs all virtualization tasks, but
more complex configurations are possible.

Adding an additional layer between the switch and control layer creates
unwanted overhead. Experiments show that response times for the processing
of Flow Requests increased from $12$ ms to $16$ ms. This means that
FlowVisor accounts for an additional delay of $4$ ms, in comparison
to non-virtualized OpenFlow topologies \cite{sherwood2009flowvisor}.
Besides the delay measurements, experiments were performed to test
bandwidth sharing between slices and CPU utilization on the hardware
OpenFlow switches. Network policies must include minimal bandwidth
guarantees (QoS parameters) for each user, to prevent unfair use of
the network.

An aspect not covered in FlowVisor is security. Additional mechanisms
for classification of traffic and checking of Flow Rules may be required
to ensure full separation and isolation between network traffic on
the slices.

\subsection{Conclusion on scalability}

Five different concepts on scalability have been reviewed. All solutions
propose to divide workload over multiple instances. However, it remains
difficult to come with an optimal scalability solution for SDN networks.
As can be seen from table \ref{tab:ComparisonSca}, trade-offs have
to be made by network designers and managers. Table \ref{tab:ComparisonSca}
gives an overview of the proposed frameworks and their components.
Besides observations from the review, also a column is defined for
standardization. This indicates availability of used components in
the framework. Unfortunately, only FlowVisor is available as open-source
software, but on conceptual level standardized components can be used
to reproduce the other proposed frameworks. A high standardization
in the table indicates that the solution is built up from standard
available OpenFlow components. A part of the table is dedicated to
data storage solutions used in ONIX and HyperFlow and is useful for
future distributed layer developments and comparisons.

\begin{table}
\protect\caption{Comparison of scalability solutions and components.}
\label{tab:ComparisonSca}

\centering
\begin{tabular}{l c c c c}

	\emph{Solution} & \emph{Standarized} & \emph{Complexity} & \emph{Decision} & \emph{Classification} \\
\toprule[2pt]

	HyperFlow & $+/-$ & $+$	& Global		& X	 \\
	ONIX 	 & $-$  & $+$    & Global		& X	 \\
	DevoFlow  & $+/-$ & $+/-$ & Semi-Global   & V 	\\
	Kandoo	& $+$    & $-$  & Semi-Global   & V	 \\
	FlowVisor & $+$ & $+/-$ & Semi-Global    & X	\\

\midrule[1pt]

    \emph{Solution} & \emph{Availability} & \emph{Performance} & \emph{Reliability} \\
\midrule[1pt]

	WheelFS      & $+/-$&  $-$  & $+$     \\
	DHT          & $+$   &  $+$    & $-$   \\
	SQL & $+$   &  $-$  & $+$     \\

\bottomrule[1.2pt] \\
\end{tabular}
\end{table}

\section{Resiliency in SDN}

\label{sec:Resiliency-in-SDN}

In regular networks, when the control logic of a switch fails, only
network traffic over that particular switch is affected. When failover
paths are preprogrammed into neighboring switches, backup paths are
available and on failure detection, backup paths can be activated.
If the control logic in an SDN enabled network fails, the forwarding
and routing capabilities of the network are down, resulting in drop
of Flow Requests, undelivered data packets and an unreliable network.
In an early stage of the development of the OpenFlow protocol, this
problem was identified and from protocol version $1.2$, a \emph{master-slave}
configuration at the control layer can be applied to increase the
network resiliency to failing OpenFlow controllers.

We define robustness of a topology as the measure of connectivity
in a network after removing links and switches. Its resiliency indicates
its ability to re-allocate redundant paths within a specified time
window. On the controller side, the robustness of a single or group
of OpenFlow controller(s) is defined as the resistance of controller(s)
before entering failure states. The resilience is defined as the ability
to recover control logic after a failure. As described, the definitions
for robustness and resiliency can have different meanings, depending
on the viewpoint of the designer. In this overview, examples and proposals
of both viewpoints are discussed.

Before proposed solutions on resiliency and robustness will be reviewed,
a small retrospect to section \ref{sec:Scalability-in-SDN} is made,
as all those solutions house the ability to increase the robustness
of an SDN network. By partitioning the workload of a single controller
over multiple instances, the robustness of the network is increased.
Failure of a single controller will only result in an uncontrollable
part of the network. To recover from a failure and increase the resiliency,
additional logic is required. For both viewpoints, timely detection
of a failure and a fast failover process are basic requirements. To
create a robust and resilient network, the network topology must include
redundant paths \cite{kuipers2012overview}. For a resilient control
layer, the network state must be synchronized and identical between
master and slave controllers. Additional modules and synchronization
schemes must meet these requirements without compromising the performance
and adding unwanted latency.

This section will review resiliency of 3 different aspects important
to the network: (1) Section \ref{sub:Replication-component} focuses
on the resiliency at the control layer; (2) Sections \ref{sub:Link-failure-recovery}
to \ref{sub:OpenFlow-based-segment} give more insight in topology
failure recovery and protection schemes, while section \ref{sub:Recovery-Inband}
discusses the more special case of in-band networks where the control
and forwarding plane share the same transport layer; (3) Finally,
section \ref{sub:securityoverview} SDN network security.

\subsection{Replication component for controller resiliency}

\label{sub:Replication-component}

\begin{figure}
\includegraphics[width=1\columnwidth]{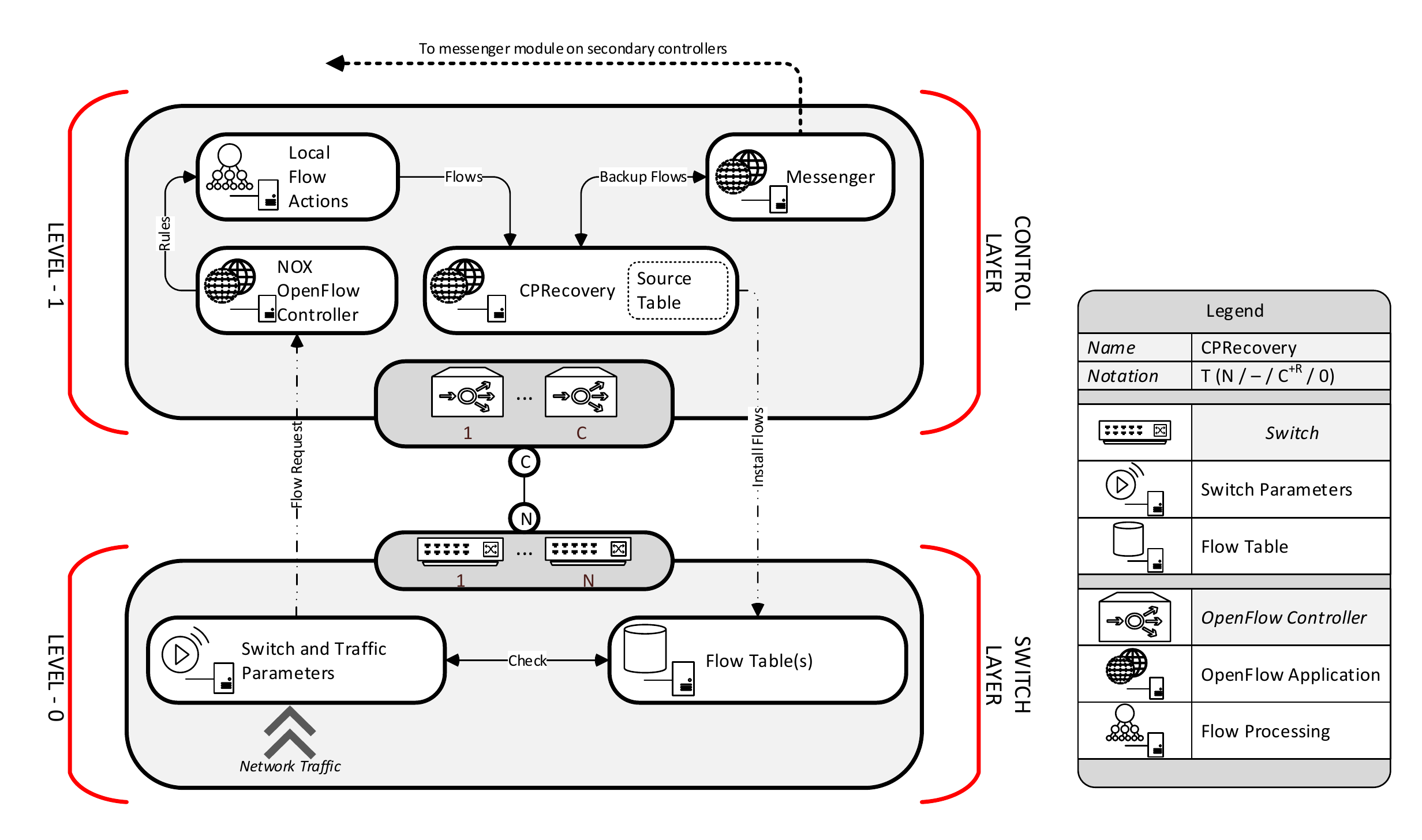}\protect\caption[Replication component in a standard NOX controller]{Replication component in a standard NOX controller - \emph{A standard
NOX-controller is enhanced to replicate flow installs over multiple
slave controllers using OpenFlow protocol 1.2 (and higher) and the
CPRecovery module.}}
\label{fig:CPR-framework}
\end{figure}

In \cite{fonseca2012replication} the master-slave capabilities of
the OpenFlow protocol are utilized to deliver controller robustness.
This indicates that a primary controller (master) has control over
all switches and on the master failure, a backup controller (slave)
can take over control of assigned switches. Fonseca et al. \cite{fonseca2012replication}
introduce a solution, indicated in this review as CPR, which integrates
a replication component into a standard OpenFlow controller. As replication
component the ``Primary-Backup'' protocol is applied, to offer resilience
against failures and a consistent view of the latest failure-free
state of the network. The primary-backup protocol synchronizes the
state of the primary controller with the backup controllers. In CPR
two phases are distinguished, namely the replication and recovery
phases. During the replication phase, calculated flows by the primary
controller are synchronized over the backup controllers. After failure
detection of the primary controller, the recovery process is initiated
to reallocate a primary controller and restore flow calculations.
With the replication component integrated in an OpenFlow controller,
the solution can be classified as a $T(N/-/C^{+R}/0)$ topology. Hereby
we denote that always one primary controller is present, with $C-1$
remaining backup controllers. The current implementation of the OpenFlow
protocol allows a total of $C=4$ controllers. In figure \ref{fig:CPR-framework}
the synchronization process of CPR is shown.

The CPR solution connects to standard OpenFlow-compliant switches
and is built upon the NOX OpenFlow controller. Additional components,
to enable replication, are integrated into the NOX controller as modules.
The switches are configured in the master-slave setting, allowing
multiple controllers in a predefined listing. During the replication
phase, flow requests are sent from the switch to the primary controller.
At the controller, the ordinary processes are executed for routing
and forwarding. After the flow is calculated in the area flow scheduler,
it is intercepted by the ``CPRecovery'' module. This module determines
whether the controller is assigned as primary and on positive identification
the flow is added to the source table of the controller. Via the ``Messenger''
module, the source table of the backup controllers are synchronized
using the primary-backup protocol. After all controllers are updated
and synchronized, the flow is installed into the switches. This replication
procedure enables a fully synchronized backup, before the flows are
installed to the switches. So when the primary fails, the second assigned
controller can seamlessly take over network control. A drawback of
this replication scheme is the additional latency introduced to the
complete flow install process.

All network switches can be configured to perform activity probing
on the primary controller. If the primary controller fails to reply
within a configurable time window ($\tau$), the network switch starts
the recovery phase and assigns the first backup controller from the
list as primary controller. On the controller side, when a join request
from the switches is received by a backup controller, this controller
will set itself as primary controller and the replication phase is
started. Update messages from the primary controller are also sent
to the original primary controller and on its recovery it is assigned
as one of the secondary controllers.

\begin{figure}
\includegraphics[width=1\columnwidth]{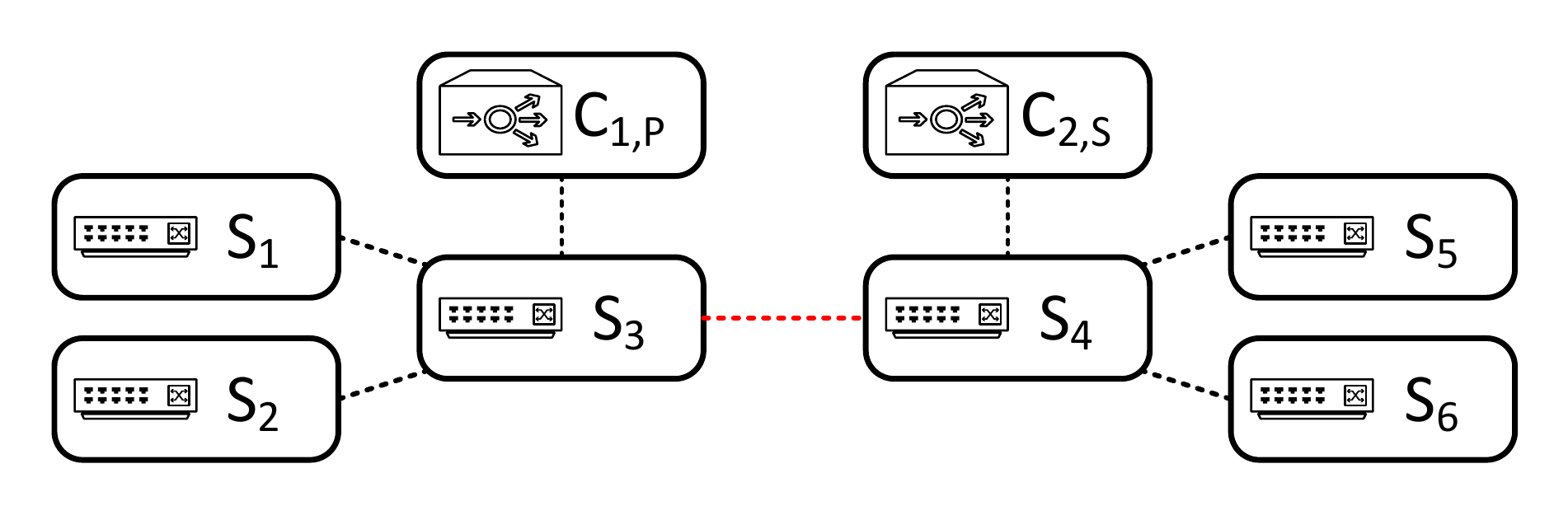}\protect\caption[Example topology for network partitioning]{Example topology for network partitioning - \emph{On a link failure
between $S_{3}$ and $S_{4}$ the topology is partitioned and the
backup controller $C_{2,S}$ will be assigned as primary by switch
$S_{4}$ to $S_{6}$ using the primary backup protocol.}}
\label{fig:CPR-example-part}

\end{figure}

The replication and recovery processes seem to solve the resiliency
problem with OpenFlow controllers, but the primary-backup protocol
and the recovery phase may fail in case of a temporary network partitioning
and geographically separated controllers. To explain the potential
flaw, the example topology $T(6/-/2^{+R}/0)$ of figure \ref{fig:CPR-example-part}
is used.

On normal operation, controller $C_{1,P}$ is assigned as primary
and controller $C_{2,S}$ as secondary (backup) controller. At time
$t$ the link $S_{3}-S_{4}$ becomes unavailable and the network is
partitioned into two components by the following reasoning. Switches
$S_{1}$ to $S_{3}$ are under control of the original primary controller,
where the remaining switches ($S_{4}$ to $S_{6}$) will select the
secondary controller as new controller, as the time window on activity
probing expires on $t+\tau$. We question the behavior of the replication
and recovery phase of the replication component (and the primary-backup
protocol) in case link $S_{3}-S_{4}$ becomes operational. Switches
$S_{4}$ to $S_{6}$ will not re-assign to the original primary controller,
so the network topology remains partitioned until failure of controller
$C_{2}$. In \cite{fonseca2012replication} and other performed research
no specific measurements are performed on the influence of geographical
positioning of OpenFlow controllers and their secondary problems,
like flow synchronizing and primary controller selection. To solve
this problem, a more advanced synchronization scheme is required,
with primary controller propagation and election schemes.

To test the functionality and the performance of the replication component,
Fonseca et al. \cite{fonseca2012replication} performed two simulations.
In the first simulation the packet delay between two hosts in a tree
topology with the primary and backup controllers connected to the
top switch are measured. At specific times the primary controller
is forced into a failure state. Where the average packet delay is
$20$ ms, the delay rises to approximately $900$ ms during the recovery
phase. After the rise, the packet delay normalizes to average and
the network functions normally. Although the replication phase is
successful, the delay during the recovery phase is unacceptable for
carrier-grade data networks providing voice services, which require
end-to-end delays not to exceed $50$ ms.

The second simulation measured the response time of a flow install.
Therefore multiple measurements have been performed using the number
of secondary controllers as variable. As described earlier, the CPRecovery
module first synchronizes the secondary controllers, before installing
a computed Flow Rule into the switch Flow Table. The measurements
show that the response time increases linearly with the number of
secondary controllers, with a minimum response time of $8$ ms when
no backup controller is configured and a maximum of $62$ ms with
3 secondary controllers to synchronize. The linear expansion of the
response times is unacceptable for data networks. We propose to increase
the performance of the CPRecovery module and lower the response times
by performing the install of the Flow Rule and the synchronization
to the secondary controllers in parallel or first installing the Flow
Rule and perform synchronization afterwards.

\subsection{Reactive link failure recovery}

\label{sub:Link-failure-recovery}

\begin{table*}
\protect\caption{Comparison of properties and results on link recovery mechanisms.}
\label{tab:LinkRes-1-Comp}\centering
\begin{tabular}{l c c c }

	\emph{Name} & \emph{Update forwarding table} & \emph{Recovery scheme} & \emph{Recovery Time} \\
\toprule[2pt]

	L2-Learning          & Traffic / ARP & - / ARP	         & Seconds - Minute  \\
	L2-Learning PySwitch & Traffic / ARP & Aging timers / ARP  & Seconds - Minute  \\
	Routing  		    & LLDP		  & Aging timers / LLDP & Seconds - Minute  \\
	Pre-determined	   & Manually      & Configured          & MilliSeconds      \\

\bottomrule[1.2pt] \\
\end{tabular}
\end{table*}

\begin{figure}
\includegraphics[width=1\columnwidth]{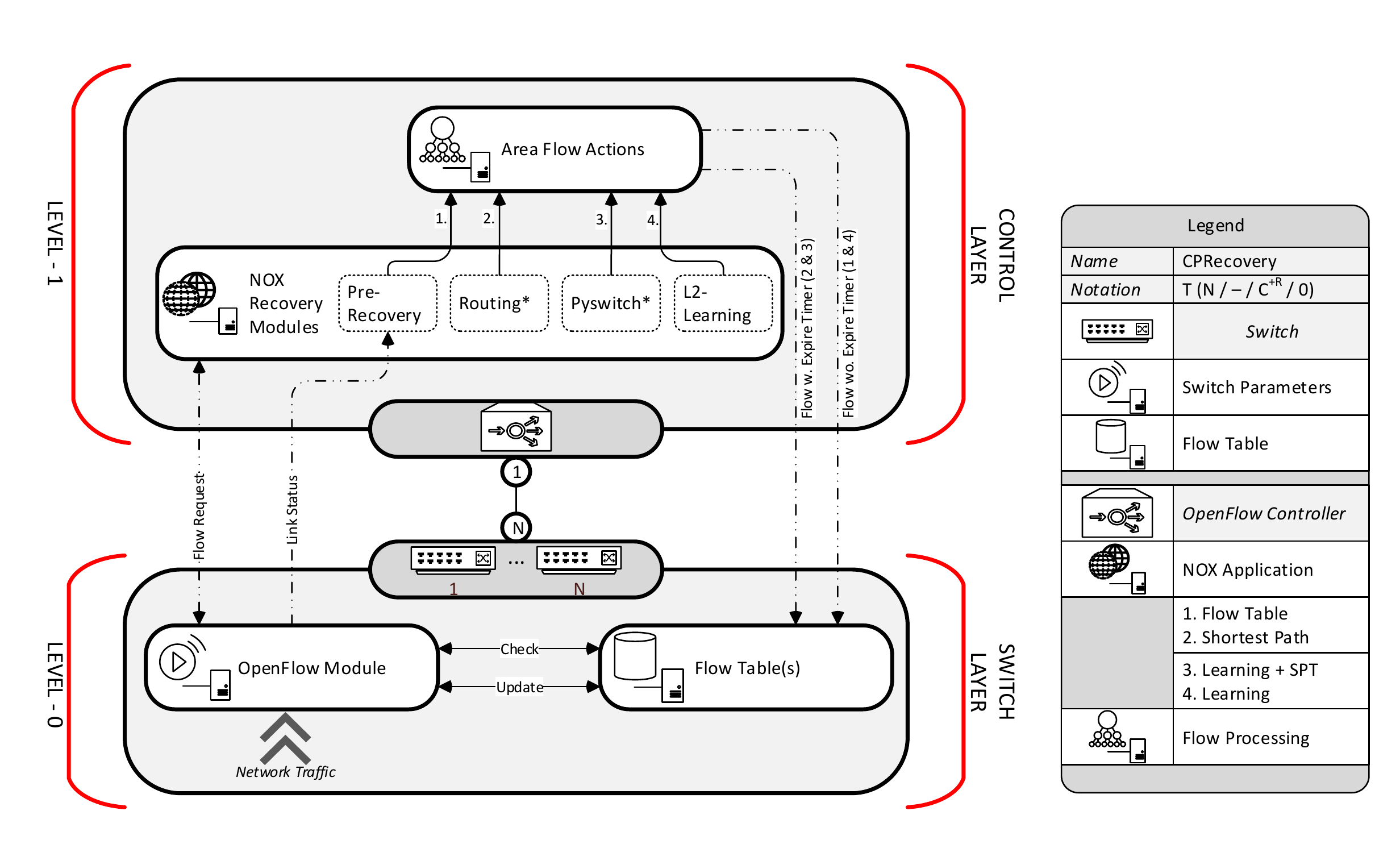}\protect\caption[Recovery mechanisms ]{Recovery mechanisms - \emph{In total four mechanisms are available
to recover from a link failure. All mechanisms have their own link
failure detection methods. On link failure, the enabled mechanism
will construct a new path and install these into the switch flow table.}}
\label{fig:Res-Link1}
\end{figure}

In \cite{sharma2011enabling} three existing switching and routing
modules (L2-Learning, PySwitch, Routing) from the NOX-controller are
compared as recovery mechanisms. Additionally, a predetermined recovery
mechanism is added. In the following the modules are discussed shortly
on their ability to recover links.
\begin{itemize}
\item \emph{L2-learning} - The standard switching module of the NOX controller
functions similarly to a common layer-2 switch. However, the applied
NOX-controller lacks the Spanning Tree Protocol (STP);
\item \emph{L2-learning PySwitch} - The functioning of this module is very
similar to the standard L2-learning module. It is extended with two
mechanisms to improve its performance. The first implemented extension
adds aging timers to the installed Flow Rules, so that the switch
can remove and update its Flow Table. Every time a switch processes
a packet, the time-stamp of the flow rule is updated. To protect the
Flow Table, the hard-time must be larger than the idle-time. The second
mechanism applied is the application of STP \cite{ieee802.1D}, to
remove possible networking loops in the topology;
\item \emph{Routing} - The routing module uses three mechanisms to compute
Flow Rules. To enable routing, the control must maintain the network
topology for path computations. To detect connected switches and link
failures, on a regular basis the switch sends Link Layer Discovery
Protocol (LLDP) packets, containing information about the switch MAC-address,
port number and VLAN indicator. A receiving OpenFlow switch replies
with an LLDP-packet containing its own parameters. When the reply
packet is received by the corresponding switch, the assigned controller
is informed about the detected link and the network topology is updated.
The recovery capabilities of the routing module depend on the discovery
mechanism and the configured timeout interval. If an LLDP-packet is
not received within the configured interval, the switch declares the
link lost and informs the controller of the status change;
\item \emph{Pre-determined} - The pre-determined module does not rely on
learning and discovery mechanisms, but it implements path protection
at the control layer. In the controller multiple static paths are
provided by the network manager. Based on priority and available paths,
the controller chooses a path and installs it on the switches accordingly.
On a network failure the controller can choose a redundant path from
the provided paths, reducing the need for link discovery mechanisms
and path calculations. As the network manager provides the paths,
no spanning tree protocol is needed (assuming that the paths are loop-free).
\end{itemize}
The first three mechanisms work dynamically and provide path restoration,
whereas the fourth mechanism is especially designed for path protection.
This module applied to the topology leads to the classification $T(N/-/C^{+R}/0)$,
because additional logic is added to the controller to improve its
performance on link resiliency. Figure \ref{fig:Res-Link1} projects
the four recovery mechanisms onto the graphical framework.

From the OpenFlow module at the switch, Flow Requests and link status
information are exchanged with the controller. In figure \ref{fig:Res-Link1}
the four recovery modules are drawn, but only one module at a time
is active. Furthermore, the routing and PySwitch modules are marked,
to indicate the availability of the spanning tree protocol. Each of
the modules can determine Flow Rules, based on the available information.
The L2-learning and pre-determined modules install Flows without aging
timers, while the routing and PySwitch modules set times to protect
flow tables at the switch. In \cite{sharma2011enabling} simulations
have been performed on a $T(6/-/1^{+R}/0)$ topology to show the behavior
of the different modules and measurements have been taken to see if
the link-recovery requirements are achievable. The topology contained
multiple cycle and Sharma et al. \cite{sharma2011enabling} showed
that much traffic is traversing between the OpenFlow switches and
the controller, to maintain link-status information. Only the pre-determined
module consumed less traffic, which is expected from its static and
fixed design. To simulate network traffic, ping packets were sent
with an interval of $10$ ms between two end-hosts. On a specified
time, a link failure was initiated and the recovery time and number
of dropped packets measured. Measurements show that it takes $108$
ms for a link failure detection to be indicated on the controller.
This value is already above the required $50$ ms, so any recovery
mechanism discussed in this research will fail. The pre-determined
module acts immediately on a link failure, which results in recovery
times of approximately $12$ ms, resulting in a total delay of $120$
ms. Results for the routing and PySwitch modules show that the recovery
depends on the idle and hard times set in the aging timer. The L2-Learning
mechanism fails recovery of a path without the application of the
Address Recovery Protocol (ARP) \cite{rfc826}. Table \ref{tab:LinkRes-1-Comp}
summarizes the properties and results of the four recovery mechanisms,
where a distinction is made on how topology (link) information is
maintained, how the mechanism recovers from link failures and on what
time scale paths are recovered.

Unfortunately, no experiments have been performed on varying the idle
and hard times of the routing and PySwitch modules. Reducing these
times, we believe, can have much influence on the recovery process
of links. Also changing the default timeout timers of ARP and LLDP
can improve the performance. The current implementation of the pre-determined
module can act fast on a link failure, but lacks the ability of constructing
paths on demand by using dynamically received topology information.

\subsection{Path-based protection}

\label{sub:Recovery-requirements}

As shown in the previous section, the reactive and pre-determined
recovery schemes implemented at the control layer do not meet the
$50$ ms recovery time requirement for carrier-grade networks. Sharma
et al. \cite{sharma2012openflow} came with a similar proposal, but
now applied on the switch layer. This reduces the recovery time, as
no communication with the control layer is required. Multiple schemes
can be applied to recover paths in case of link failures, where a
distinction can be made between protection and restoration schemes
\cite{kuipers2012overview}. Protection schemes do not need communication
with the controller to restore paths, as actions are pre-configured
at the switch layer. Restoration schemes require communication between
the switch and controller and recovery paths are dynamically allocated. 

In \cite{sharma2012openflow} the $1:1$ protection scheme is implemented
as a protection mechanism at the switch layer, where $1:1$ refers
to activating a backup path after failure of the primary path. To
enable this mechanism, the Group Table concept of the OpenFlow protocol
is utilized. In normal operation, the destination address from a packet
is matched in the Flow Table and the packet will be forwarded to the
correct port or dropped. By applying Group Tables, a flow rule can
also contain a link to a unique group. In the Group Table, one or
more Action Buckets define actions based on status parameters. On
change of these parameters, the action bucket executes a predefined
action. In case of the protection scheme, when a failure is detected
in the path, the backup path is enabled for the flow. For path failure
detection the Bidirectional Forwarding Detection (BFD) \cite{rfc5880}
protocol is implemented. To monitor the complete path between multiple
OpenFlow switches, a BFD session is configured between the entry and
exit switches. If the periodical messaging over the session fails,
BFD assumes the path lost, updates the action bucket in the OpenFlow
switches and the protected path is installed. The $1:1$ protection
scheme implemented on a topology leads to a $T(N^{+R}/-/1/0)$ system.

\begin{figure}
\includegraphics[width=1\columnwidth]{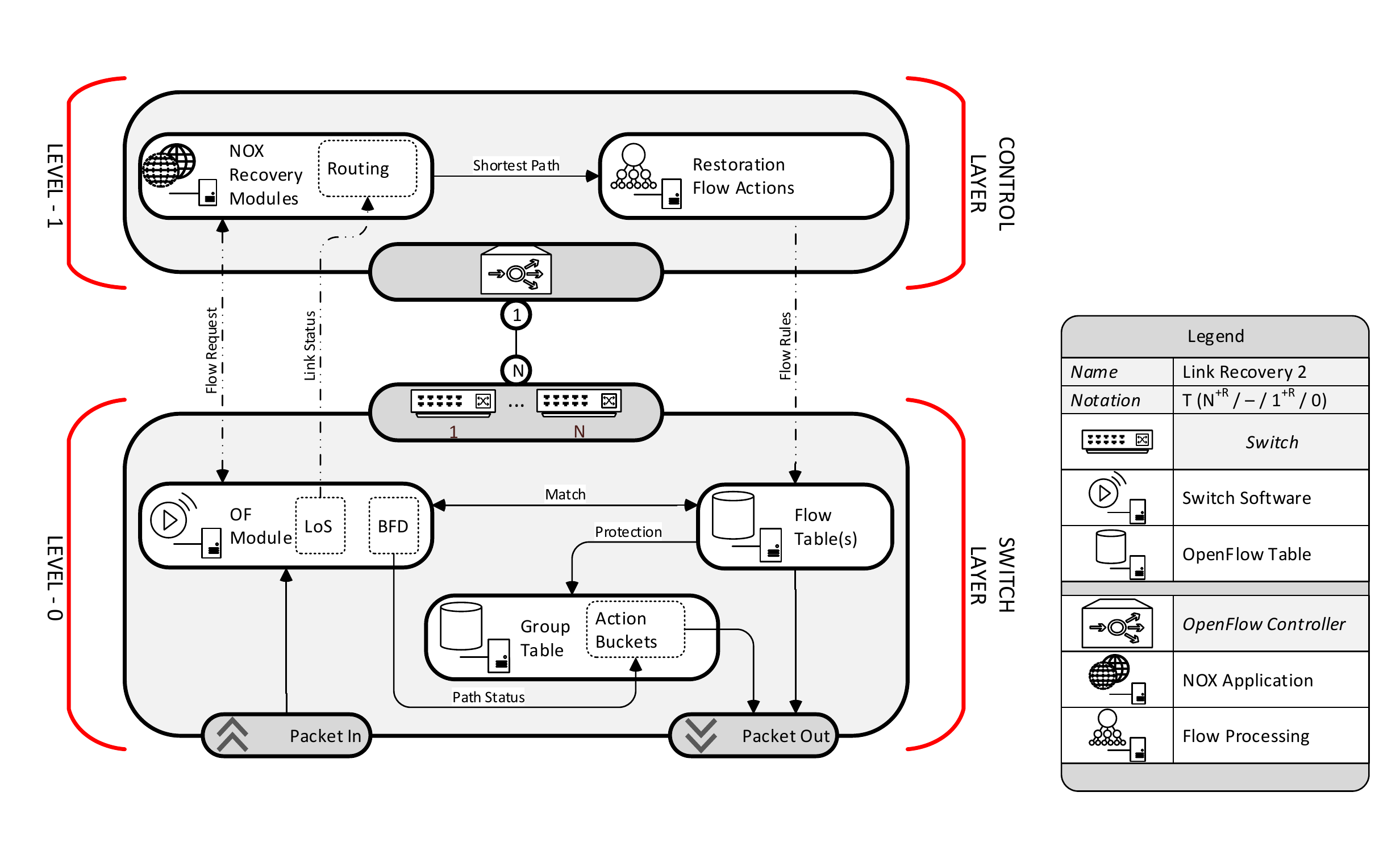}\protect\caption[$1:1$ Recovery scheme ]{$1:1$ Recovery scheme - \emph{Two schemes are visible to recover
from a failure. The protection scheme utilizes the BFD path failure
detection protocol in cooperation with Group Tables and Action Buckets
to enable $1:1$ path protection. Restoration of failed links is executed
by the controller and a modified routing module which uses LoS for
link failure detection. New constructed paths are installed to the
switches, without incorporating the failed link.}}
\label{fig:Res-Link2}
\end{figure}

The second implementation in \cite{sharma2012openflow} is the $1:1$
restoration scheme at the switch layer. An extension to the standard
routing module of the NOX controller is made to increase the resiliency.
The failure detection capabilities of the routing module depend on
the OpenFlow aging timers and the implementation of a topology module
incorporating LLDP packets. The extended routing module uses the ``Loss
of Signal'' (LoS) failure detection mechanism available in the OpenFlow
protocol. LoS detects port changes in the switch from ``Up'' to
``Down'' and reports these to the controller. Other than BFD, LoS
does not monitor complete paths, but only local links at the switch.
On link failure detection, a notification is sent to the routing module
and a new path is constructed, without incorporating the failed link.
The new path with its corresponding flow rules are installed in the
switches, after which the path is recovered. The proposed solution
for path restoration is classified as $T(N/-/1^{+R}/0)$.

The protection scheme most likely restores paths faster, as no communication
is required with the controller and backup paths are preconfigured.
The restoration scheme is more adaptive and is more flexible, as paths
are calculated with status parameters of the current topology. In
a large network, both schemes can be applied, depending on the network
services provided. A combined scheme can be classified as a $T(N^{+R}/-/1^{+R}/0)$
topology, as shown in figure \ref{fig:Res-Link2}.

In normal operation, the process of packet forwarding is similar to
the standard OpenFlow operation. On protected paths, the BFD protocol
monitors the status and on failure the action buckets in the Group
Table are updated. Actions defined in the Action Buckets, enable the
protected path. In case of restoration, the OpenFlow module monitors
a link failure, after which the routing module in the controller constructs
and installs a new path. An important aspect of the recovery process
is the latency between time of link failure and the recovery of all
affected flows. In \cite{sharma2012openflow} an analytical model
is given for the restoration process. It gives a good indication where
the latency is introduced in the recovery process. We have extended
the model with the protection scheme, to indicate the differences
between both recovery schemes. 

\begin{equation}
T_{R}=T_{LoS}+\sum_{i=1}^{F}(T_{LU,i}+T_{C,i}+T_{I,i})\label{eq:AnModRestoration}
\end{equation}

\begin{eqnarray}
T_{P} & = & max(T_{BFD,1},...,T_{BFD,N})+\label{eq:AnModProtection}\\
 &  & \sum_{i=1}^{P}max(T_{AB,1,i},...,T_{AB,N,i})\nonumber 
\end{eqnarray}

The total restoration time ($T_{R}$) is determined by the loss-of-signal
failure detection time ($T_{LoS}$), the total time spent at the controller
to look up the failed link ($T_{LU}$), the path calculation time
($T_{CALC}$), the flow install / modification time ($T_{I}$) and
the number of flows ($F$) to restore. In here the propagation delay,
which is assumed to be small ($\sim1$ms), is integrated with the
failure detection and flow installation time. The protection model
depends on the BFD failure detection time ($T_{BFD}$), the time to
process the action bucket ($T_{AB}$) and the number of flows affected
by the link failure ($P$). Because a broken flow is only restored
after the processing of the ``slowest'' of $N$ switches in the
path, the max operator is applied. 

To give an indication of the latency differences, multiple simulations
and measurements have been performed on different topologies in \cite{sharma2012openflow}.
Results are show in table \ref{tab:Link2-Time}. Delay times to process
the action buckets are unknown and likely not more than several milliseconds.

\begin{table*}
\protect\caption{Comparison of time delays in link restoration and protection.}
\label{tab:Link2-Time}\centering
\begin{tabular}{l c c l l}

	\emph{Time} & \emph{Symbol} & \emph{Delay (ms)} & \emph{Relation} & \emph{Comment} \\
\toprule[2pt]

	Failure detection time ($P$)   & $T_{BFD}$ & 40 - 44   & Fixed  &  \\

\midrule[0.5pt]
	
	Failure detection time ($R$)    & $T_{LoS}$ & 100 - 200 & Fixed  &  \\
	Controller look-up time ($R$)   & $T_{LU}$     & 1 - 10    & Linear & Delay with 250 - 3000 flows  \\
    Path calculation time ($R$)	 & $T_{CALC}$   & 10 - 200  & Linear & Delay with 25 - 300 paths      \\
    Flow installation time ($R$)    & $T_{I}$      & 1 - 5     & Linear & Delay with 1000 - 10000 flows  \\

\bottomrule[1.2pt]\\
\end{tabular}
\end{table*}

Both recovery schemes were able to recover paths on link failure.
The main difference in performance is found in the failure detection
mechanism. Where BFD only needs $40$ ms to detect a path failure,
the LoS mechanism takes more than 100 ms to report a broken link.
A main disadvantage of the application of BFD in the protection scheme
is the introduced overhead for monitoring all paths. Furthermore,
the fixed pre-planned configuration is inflexible and the experiments
were performed in such a way that link failures did not influence
the protected paths. Restoration is more flexible by allocating restoration
paths dynamically with up-to-date network topology information. Recovery
times for both schemes mainly depend on the number of flows to recover
in the network. Restoration times can exceed $1000$ ms, if a large
number of flows need recovering.

\subsection{Link-based protection}

\label{sub:Openflow-Link-based-protection}

In addition to the path-based discovery discussed in the previous
section, Van Adrichem et al. \cite{vanadrichemopenflowrecovery} propose
to deploy link-based monitoring and protection to overcome topology
failure. Their contribution in minimizing recovery time is twofold:
\begin{enumerate}
\item They minimize failure detection time. By using link-based, instead
of path-based, BFD monitoring sessions, the per-session RTT and thus
BFD interval window is minimized compared to per-path sessions. Experiments
show that configurations with a BFD interval window of $1$ ms are
feasible.
\item They adapt the Group Table implementation of the OpenFlow capable
software switch \emph{Open vSwitch} to consider BFD status real-time,
hence eliminating the administrative processes of bringing an interface'
status down.
\end{enumerate}
Herewith, they enable a controller to employ protection by configuring
per-switch backup paths using BFD aware Group Table rules. Where path-based
failure monitoring has a complexity of $O(N\times N)$ sessions, link-based
failure monitoring decreases to a complexity of $O(L)$, where $N$
and $L$ respectively represent the number of nodes and links in a
network. Hence, the number of BFD sessions traversing each link is
limited to exactly $1$.

The experiments of \cite{vanadrichemopenflowrecovery} show a recovery
time as low as $3.3$ ms independent of network size. Instead, due
to the software nature of Open vSwitch - the solution scales to the
degree of each node, emphasizing the need for hardware implementations
of BFD and packet forwarding.

Since the proposed solution deploys per-switch backup paths, in exceptional
cases crankback routing may need to be applied. Where a fast failure
recovery is still guaranteed, the solution is suboptimal. However,
in time the network controller will be notified of the change in topology
and can reconfigure the network to an optimal state without service
interruption.

\subsection{Segment-based protection}

\label{sub:OpenFlow-based-segment}

\begin{figure}
\includegraphics[width=1\columnwidth]{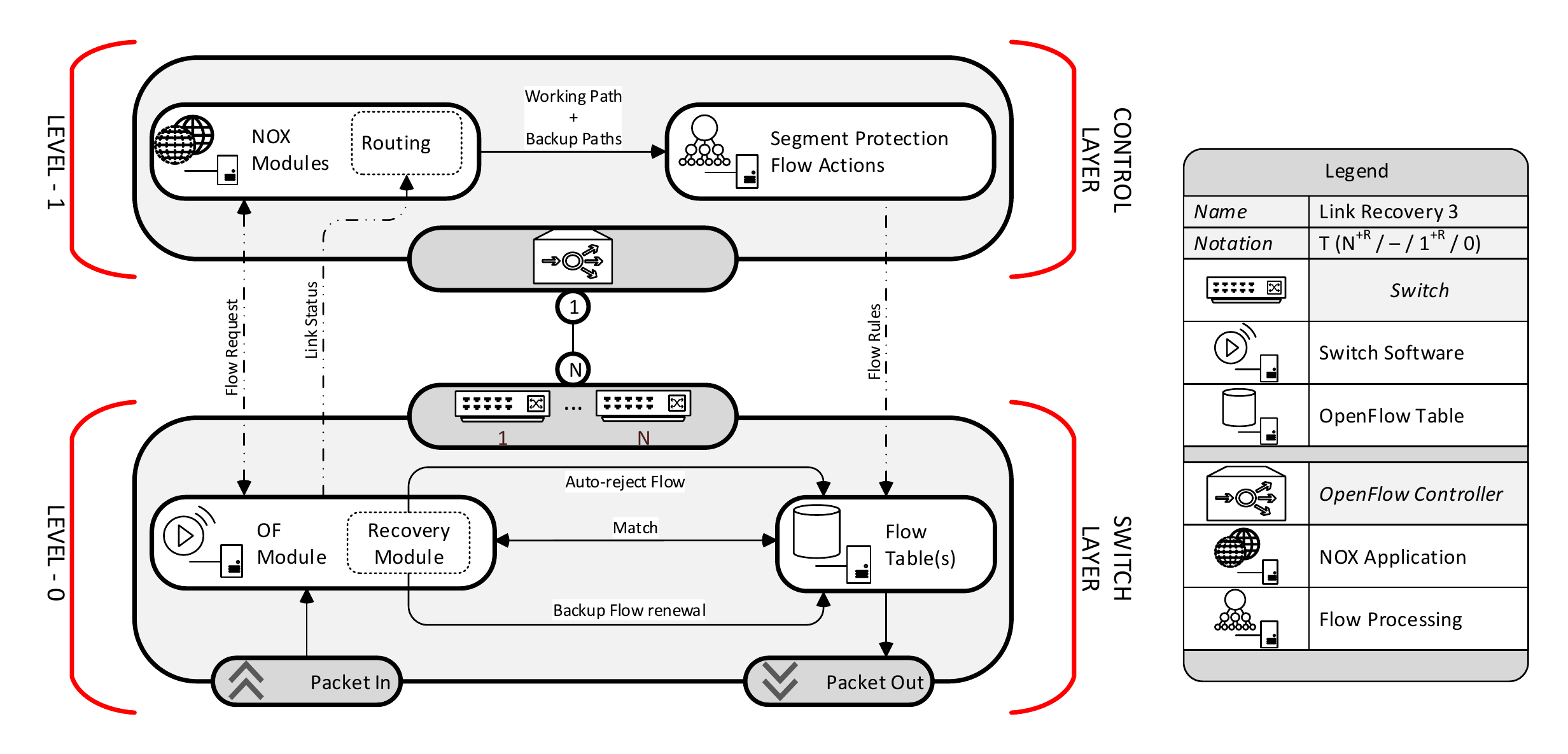}\protect\caption[OpenFlow segment protection scheme]{ OpenFlow segment protection scheme - \emph{Along with the working
path, backup paths are provided to the OpenFlow switch to protect
segments. An extended Recovery Module in the OpenFlow module rejects
flows from the Flow Table after a failure detection and enables backup
paths for the segments. To prevent backup flows from removal by the
idle timers, the recovery module transmits renewal messages over the
backup paths.}}
\label{fig:Res-Link3}
\end{figure}

Where the previous two sections discuss path- and link-based protection
against network failure, the research in \cite{sgambelluri2013openflow}
proposes a hybrid approach, where individual segments of a path are
protected. The main idea is to provide a working path, as well as
a backup path for each switch invoked in the working path. Both paths
are installed in the Flow Tables with different priorities and after
failure detection, the flows for the working path are removed from
the table by additional mechanisms in OpenFlow, after which the backup
flow becomes the working path. In figure \ref{fig:Res-Link3} the
projection to the graphical framework is given (note the overlap with
figure \ref{fig:Res-Link2}).

The failure detection mechanism in \cite{sgambelluri2013openflow}
is unknown and to trigger the backup paths, two additional modules
are added to the OpenFlow protocol version 1.0. For removing the flows
for the working path from the Flow Table, an ``auto-reject'' mechanism
is developed. This mechanism deletes entries when the port status
from the OpenFlow switches changes. The second developed mechanism
``flow renewal'' is used to update flow entries for the backup paths.
Backup paths are installed using idle timers and while the working
path is active, update messages are transmitted over the backup paths
to update the idle timers, preventing automatic flow removal by OpenFlow.

Multiple experiments have been performed with the adapted version
of OpenFlow, utilizing segment protection with the ``auto-reject''
and ``flow renewal'' mechanisms. Results show average recovery times
with a variable number of flow entries per switch around $30$ ms,
with a maximum of $65$ ms. The fact that modifications and extensions
must be made to the OpenFlow protocol, leading to a non-standard implementation,
makes that we do not recommend the solution in \cite{sgambelluri2013openflow}
for large SDN implementations.

\subsection{In-band OpenFlow networks}

\label{sub:Recovery-Inband}

In sections (\ref{sub:Link-failure-recovery}) to (\ref{sub:OpenFlow-based-segment}),
the controller was connected in an ``out-of-band'' configuration,
which indicates that separate connections from the switch to the controller
are available. Only control traffic traverses over these connections,
ensuring no delays or traffic congestion between controller and switches.
In an ``in-band'' configuration, control traffic traverses the same
connections as data traffic. No additional network interfaces are
needed. With the application of an in-band configuration to a network,
Sharma et al. \cite{sharma2013fast} discovered a problem. When a
link failure occurs and the communication between a switch and controller
is lost, the basic operations for a switch are to restore the connection
by requesting a new connection after waiting for an echo request timeout.
The minimum value for the timeout is limited to $1$ second, which
is a magnitude of $20$ too long for proper path recovery in carrier-grade
networks. Therefore in \cite{sharma2013fast} the restoration and
protection schemes from \cite{sharma2012openflow} are reused to solve
the problem. 

\begin{figure}

\centering{}\includegraphics[width=0.5\columnwidth]{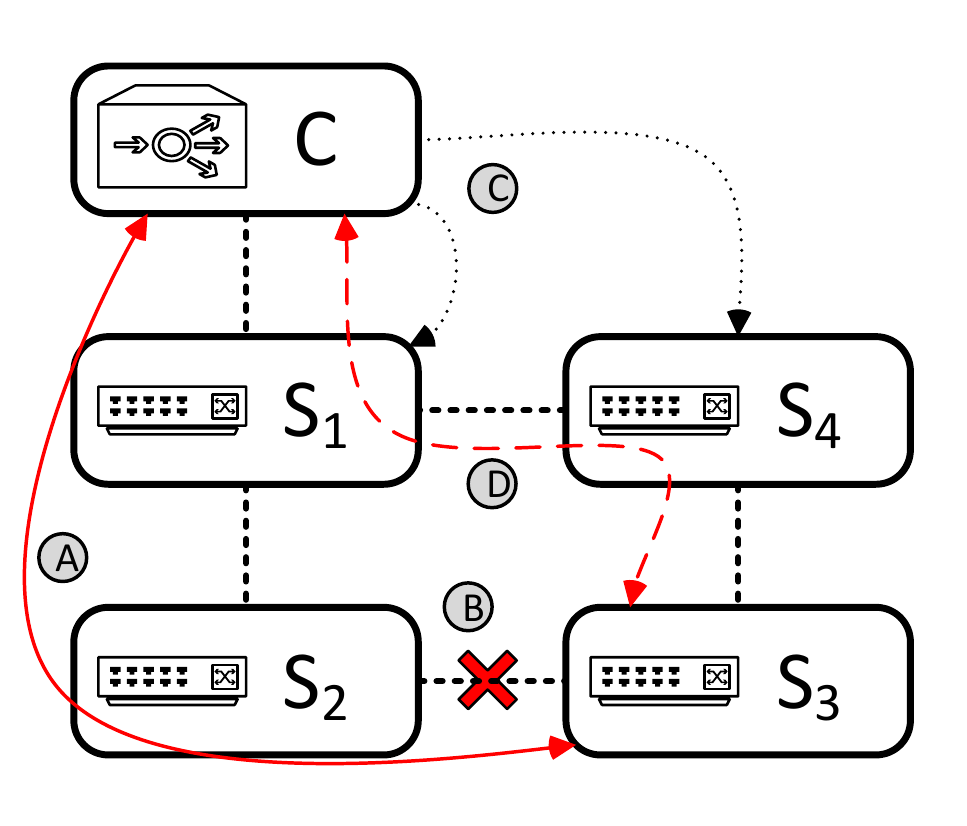}\protect\caption[Example configuration for control traffic restoration]{Example configuration for control traffic restoration -\emph{ A.
Normal situation. B. Link failure and controller notification. C.
Update intermediate switches to restore control path. D. Communication
between $S3$ and controller is restored.}}
\label{fig:ExampleLinkRes3}
\end{figure}

As seen before, data traffic paths can be restored or protected from
a link failure. In case of restoration, the failed data traffic paths
cannot be restored without communication channels between the switches
and the controller. Therefore, first the control path must be restored
after which the data paths can be reinstalled on the switches by the
controller. In order to implement this priority of processing the
link failure, the ``Barrier Request and Reply Messages'' concept
from the OpenFlow protocol is utilized. In ``normal'' operation,
OpenFlow messages can be reordered by the switches for performance
gains. To stop reordering, Barrier Messages are sent by the controller
and the switches must, upon receiving a Barrier Request, process all
preceding instructions before processing the following request. After
processing of the Barrier Request, a reply is sent to the controller.
To clarify the restoration process in an in-band configuration, an
example topology with process description is given in figure \ref{fig:ExampleLinkRes3}.

In figure \ref{fig:ExampleLinkRes3}, in total four phases are distinguished
from normal operation to link failure and restoration of the control
channel for switch $S_{3}$:
\begin{itemize}
\item \emph{Phase A} - Initial phase where the control traffic for switch
$S_{3}$ is routed over switch $S_{1}$ and $S_{2}$ to the controller.
In a normal ``out-of-band'' configuration, $S_{3}$ would have a separate
network connection with the controller;
\item \emph{Phase B} - The link between switches $S_{2}$ and $S_{3}$ fails
and the communication between $S_{3}$ and the controller stops. Switch
$S_{2}$ monitors the link failure with the LoS mechanism and sends
an error report to the controller via $S_{1}$;
\item \emph{Phase C} - The controller calculates the new control path over
$S_{1}$ and $S_{4}$ to $S_{3}$ with highest priority, whereafter
the data traffic paths are recalculated. These paths cannot yet be
installed into switch $S_{3}$, as the broken path is still present
in the flow table. Therefore the controller first updates $S_{1}$
and $S_{4}$;
\item \emph{Phase D} - The flow modification messages are processed and
the reply message is sent to the controller by $S_{1}$ and $S_{4}$.
After both barrier reply messages are received by the controller,
the new control path to switch $S_{3}$ is configured at the intermediate
switches. The connection to the controller is restored and the recalculated
data paths can be installed in all switches.
\end{itemize}
As seen in the description of the phases, the use of barrier requests
synchronizes the intermediate steps of restoration. Besides restoration,
the control traffic path can also be recovered by a $1:1$ protection
scheme. Protection of the control and data traffic paths is provided
by BFD link failure detection and the Group Tables with Action Buckets.
Main advantages of protection is that no communication is required
and the switches can autonomously update their flow tables. 

Using the restoration and protection schemes, a total of four recovery
schemes are possible. As with the out-of-band configuration, analytical
models can be used to predict the behavior during the recovery process.
The restoration model for the control traffic path is a modification
of the earlier restoration model. Equations (\ref{eq:AnModConRestoration})
and (\ref{eq:AnModDataRestoration}) show the restoration times for
an OpenFlow in-band configuration, where $T_{RC}$ is the control
traffic restoration time, $T_{B}$ is the additional time delay introduced
by the Barrier Message reply mechanism and $T_{IS,i}$ is the time
to install and modify the flow tables of intermediate switches. The
restoration time for data traffic paths ($T_{RD}$) is a simplified
form of equation (\ref{eq:AnModRestoration}), without the LoS failure
detection delay as the controller already is informed about the network
failure.

\begin{equation}
T_{RC}=T_{LoS}+T_{B}+\sum_{i=1}^{S}(T_{LU,i}+T_{C,i}+T_{IS,i}+T_{I,i})\label{eq:AnModConRestoration}
\end{equation}

\begin{equation}
T_{RD}=\sum_{i=1}^{F}(T_{LU,i}+T_{C,i}+T_{I,i})\label{eq:AnModDataRestoration}
\end{equation}
In table \ref{tab: Link3-TimeOverview} the four recovery schemes
are given, together with the analytical delay models.

\begin{table}
\protect\caption{Comparison of recovery schemes in in-band configurations.}
\label{tab: Link3-TimeOverview}\centering
\begin{tabular}{l c l}

	\emph{Recovery Scheme (Control - Data)} & \emph{Symbol} & \emph{Analytical Relationship}  \\
\toprule[2pt]

	Restoration - Restoration & $T_{R,R}$ & $T_{RC}+T_{RD}$     \\

\midrule[0.3pt]

    Restoration - Protection  &$T_{R,R}$  & $max(T_{RC},T_{P})$   \\

\midrule[0.3pt]

    Protection - Protection   &$T_{P,P}$  & $T_{P}$ \\

\midrule[0.3pt]

    Protection - Restoration  &$T_{P,R}$  & $max(T_{P},T_{R})$ \\

\bottomrule[1.2pt]\\
\end{tabular}

\end{table}

The analytical relationships in table \ref{tab: Link3-TimeOverview}
assume that the recovery and protection processes do not influence
each other at the switch. In \cite{sharma2013fast} multiple measurements
have been preformed on all four in-band recovery schemes. Results
show that when restoration is applied to recover data paths, delays
exceed the $50$ ms requirement. Only the full protection recovery
scheme meets the requirements, but in practice this scheme will not
be applied due to large flow tables and the large number of configurations
which have to be made by the network manager at the switches. 

Looking to the differences in the results between $T_{R,R}$ and $T_{P,R}$,
we can conclude that the performance difference is small%
\footnote{Results show a $T_{LoS}$ of approximately $50$ ms in comparison
with $100$ ms measured in \cite{sharma2012openflow}.%
}. This is expected, as only a few control paths to the switches have
to be recovered. With this conclusion, we can state that for recovery
requirements there is no noticeable performance difference between
in- and out-of-band configurations, when a full protection scheme
is implemented. A comparison between restoration in both configurations
is not possible due to the differentiation in delay in the measurements
of \cite{sharma2012openflow} and \cite{sharma2013fast}. We think
it is possible to predict the behavior for both configurations with
the derived analytical models, when reliable measurements for the
defined parameters are present.

\subsection{Security in SDN}

\label{sub:securityoverview}

\begin{figure}
\includegraphics[width=1\columnwidth]{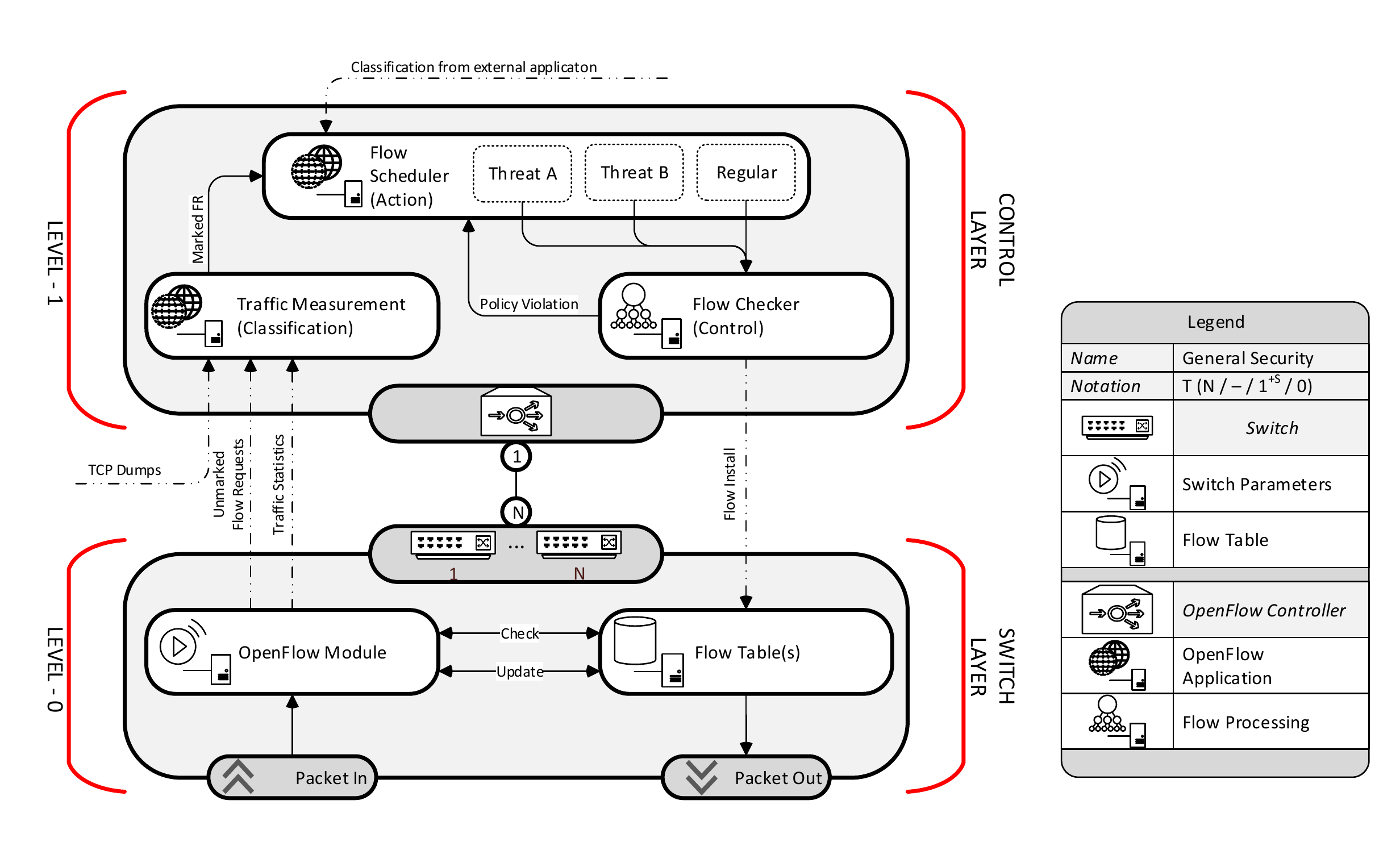}\protect\caption[General security configuration]{General security configuration - \emph{The first step is to classify
and mark incoming traffic with statistics from the OpenFlow switch
or external measurements (Classification). Marked Flow Requests go
to the Flow Scheduler where the requests are processed and according
actions are assigned to flows (Action). The last step is to check
the flows to the network security policy (Control) and install the
flows at the switches. On policy violation the Flow Scheduler must
be informed. }}
\label{fig:SDNR-SecurityFramework}
\end{figure}

Network security is applied to control networks access, provide separation
between users and protect the network against malicious and unwanted
intruders. It remains a hot topic under SDN researchers, because a
basic security level is expected from a new network technology, as
well as the fact that network security applications can easily be
applied to the network control logic. We define two levels of security.
The first level invokes logical connections between end hosts inside
the network. Protocols like Secure Socket Layer (SSL) or packet encrypting
techniques must ensure connection security. Within SDN, this level
of security plays an important role, as the control link between switches
and the centralized controller must be ensured. In the OpenFlow protocol
a mechanism to secure the connection is available, but not required.
It is up to the controller to secure the connection with the switches
and a number of controller implementations have not implemented link
security mechanisms. When no link security is applied, a malicious
node can impersonate the controller and take over control of the switches. 

The second level of security is to protect switches, servers and end
hosts in the network. Numerous examples are present to indicate the
threats to the network as a whole. Malicious software can intrude
the network, infect hosts and gather information, but also flooding
attacks can disable network servers or overload OpenFlow switches
and controllers. Security mechanisms must be implemented on the network
to detect malicious traffic and take necessary actions to block and
reroute this traffic. In current networking, network security is applied
at higher networking layers. Routers and firewalls perform security
tasks at layer 3, whereas end hosts and servers host security applications
at layer 7. With SDN, there is a central authority that routes traffic
through the network and enables the possibility to apply security
policies to all layers in networking. Much research has been performed
and the results of \cite{shin2013fresco,braga2010lightweight,sonmodel,porras2012security}
are used to determine security properties within SDN. Most researchers
follow roughly the same procedure to apply security to the network.
This procedure consists on three steps where a short description of
the process is given, as well as a reference to the performed research.
\begin{itemize}
\item \emph{Classification} - Data flows through the network must be classified
in order to determine malicious behavior and network attacks. Without
classification it is impossible to protect the network and perform
countermeasures. The main source for traffic classification is found
in traffic statistics \cite{braga2010lightweight};
\item \emph{Action }- Once a traffic flow is marked as malicious, the control
layer must modify flow tables to protect the network and prevent propagation
of the malicious traffic through the network. For each threat, different
actions are needed, so the control layer must be flexible for quick
adoption of new protection schemes \cite{shin2013fresco}; 
\item \emph{Check }- The last security process is the checking of calculated
flow rules with the applied security policy from the network manager.
Flow rules may (unintentionally) disrupt the security policy and therefore
an extra control process is needed. Preventing network security violations
by checking flow rules before install on the switches, completes the
overall security process \cite{sonmodel,porras2012security}.
\end{itemize}
The three processes combined form the protection layer for the network
and all can be implemented at the control layer, which results in
a $T(N/-/1^{+S}/0)$ configuration. To give the most general view
of this configuration, we assume no modifications to the switch layer.
Figure \ref{fig:SDNR-SecurityFramework} gives the general OpenFlow
security configuration.

As seen in figure \ref{fig:SDNR-SecurityFramework}, normal Flow Requests
and traffic statistics enter the classification module in the OpenFlow
controller. Traffic statistics can originate from the OpenFlow module
at the switch and result from processed TCP and UDP traffic dumps.
The classification module identifies malicious traffic flows and has
two actions to perform. First, it must inform the Flow Scheduler with
the presence of malicious flows in the network. Existing flows in
the switch tables must be modified by the Flow Scheduler. Second,
incoming Flow Requests must be marked, so that the flow scheduler
can process the Flow Requests according to the security policy. At
the scheduler, multiple security modules are present, to install and
modify Flow Tables with rules based on the security policy for regular
traffic and counter measures for the different threats to the network.
So for each traffic flow there exists a unique set of rules, in order
to protect the individual traffic flows within the network, as well
as the network itself. The last step before Flow Rules can be installed
or modified is confirming validity to the overall security policy
of the network. A computed Flow Rule by a module in the Flow Scheduler
can confirm the rule of that module, but may violate the security
policies of other modules and the overall network security policy.
After a Flow Rule is approved by such a flow policy checker, it can
be installed into the switch Flow Tables.

In theory, the classification process looks easy to execute, but \cite{braga2010lightweight}
and \cite{shin2013fresco} have proven otherwise. In \cite{braga2010lightweight}
an effective solution is found to identify abnormal traffic and flooding
attacks. The most obvious mechanism to classify traffic flows are
continuous TCP and UDP traffic dumps. With these dumps, all information
is present to identify malicious traffic, but it takes much computing
resources to process all the dumps continuously. Therefore an intelligent
mechanism is employed to map traffic flows based on traffic statistics.
Using the maps, all flows are characterized and abnormal traffic flows
can be identified and removed from the network. This method is only
to detect flooding attacks, so to detect other threats, more classification
mechanisms are needed. In \cite{shin2013fresco} an application layer
is presented to apply classification modules, as well as security
modules, at the Flow Scheduler. A general application layer eases
implementation of modules for newly identified threats. Multiple examples
in \cite{shin2013fresco} show that with the application of classification
and security modules, counter measures for network threats can be
implemented into an OpenFlow environment.

\subsection{Conclusion on resiliency}

\label{sub:Conclusion-on-resiliency}

Where \cite{fonseca2012replication} focuses on the resiliency of
the control plane and developed a mechanism to utilize the master-slave
concept for OpenFlow controllers, \cite{sharma2011enabling,sharma2012openflow,vanadrichemopenflowrecovery,sgambelluri2013openflow,sharma2013fast}
research the ability to recover failed links. On the controller side
we can state that a single controller is insufficient in terms of
robustness by a lack of redundancy. With the failure of a controller
the OpenFlow switches will lose the control layer functionality, resulting
in an uncontrolled network topology. If the $50$ ms recovery requirement
from carrier-grade networks is applied, the replication scheme used
to improve robustness will not suffice. Hence, the replication component
needs modifications to lower the repair latency.

On link recovery, five papers are discussed. Although most recovery
concepts start with ``traditional'' failure recovery based on Loss-of-Signal
to detect failure, these methods have proven to have a large latency.
Instead, actively probing failure detection mechanisms can detect
failures more quickly. Group Tables with pre-programmed failover logic
and fast restoration of flows by the controller are two proven techniques
which have shown to recover paths within sub $50$ ms time. Hence,
recovery from topology failure appears sufficiently covered.

However, an integral solution, where also OpenFlow controllers are
redundantly applied, with the ability to recover network control in
millisecond order, needs further research. Techniques used in the
scalability research field can be joined with the OpenFlow protocol
to apply master-slave configurations. Combined with proposed path
recovery schemes a higher level of resilience can be reached in both
the control and forwarding plane.

\section{Conclusion}

\label{sec:Conclusion}

In this overview, we have discussed the basic principles of SDN, where
the control layer is decoupled from the data plane and merged into
a centralized control logic. The centralized logic, itself controlled
by software, has a global view of the network and has the capabilities
to dynamically control hardware devices for optimal traffic flows
through the network. For communication between the data plane and
the control logic, the OpenFlow protocol is commonly utilized. Two
main problem areas are identified from the reviewed research, being
limited scalability and decreased resiliency due to the centralized
nature of SDN. 

To gain detailed insight on performed research in SDN and OpenFlow
networks, we have developed a general framework and notation in which
we classify scientific work related to scalability and resiliency
in SDN. We have made a separation in proposed solutions based on scalability
and resiliency issues. Due to the centralization of the control logic,
scalability issues exist on the number of hardware devices to control
by a single control logic or the number of Flow Requests processed
by the logic. Solutions can be found in increasing performance of
the central logic, reducing the number of tasks to perform by the
central control logic or dividing hardware resources with either virtualization
or introducing multiple coexisting controllers.

On resiliency three problem areas have been distinguished, being the
resistance of both the controller and the network topology against
failures, as well as the network resilience against malicious attacks
(security). Resiliency of SDN networks can be increased with the application
of redundant network controllers and replication of internal network
state, while recovery schemes can protect against network topology
failures.

However, as each implementation seems to make trade-offs, possible
solutions are still suboptimal by nature. The topics of scalability
and topology failure may be sufficiently solved by combining the decrease
of control overhead, distributing multiple controllers among a network,
and deploying discussed failure protection mechanisms. The topic of
recovery from controller failure, however, seems underrepresented
and needs to be researched more thoroughly.

\bibliographystyle{IEEEtran}
\bibliography{References,rfc}

\end{document}